\documentclass[12pt, amsfonts, epsfig, reqno, a4paper]{amsart}
\usepackage[hyperindex,breaklinks]{hyperref}
\usepackage{amsmath}
\usepackage{amsfonts}
\usepackage{mathrsfs}
\usepackage{graphicx}
\usepackage{color}
\usepackage{slashed}
\usepackage{braket}
\usepackage{extarrows}
\usepackage{amsmath, bm}
\usepackage{amssymb}
\usepackage[margin=1in]{geometry}
\usepackage{bbm}
\usepackage{hyperref}

\newcommand{\half}{\frac12}

\definecolor{yellow}{rgb}{1,.7,0}

\definecolor{pkured}{rgb}{0.55,0,0}

\newcommand{\be}{\begin{equation*}}
	\newcommand{\ee}{\end{equation*}}
\newcommand{\beq}{\begin{equation}}
	\newcommand{\eeq}{\end{equation}}

\numberwithin{equation}{section}

\newtheorem{cor}{Corollary}[section]

\newtheorem{lem}[cor]{Lemma}
\newtheorem{prop}[cor]{Proposition}
\newtheorem{thm}[cor]{Theorem}

\theoremstyle{remark}
\newtheorem{rmk}[cor]{Remark}

\numberwithin{figure}{section}
\usepackage{ytableau}

\usepackage{tikz}
\newcounter{x}
\newcounter{y}
\newcounter{z}

\author{Chenglang Yang}
\email{yangcl@pku.edu.cn}
\address{Institute for Math and AI, Wuhan University, Wuhan 430072, China}

\title[{On Two Families of Nekrasov--Okounkov Type Formulas}]
{On Two Families of Nekrasov--Okounkov Type Formulas}

\begin{document}
	\maketitle

	\begin{abstract}
		In this paper,
		we use the vacuum expectation value formula for the topological vertex and its rotation symmetry to derive two families of Nekrasov--Okounkov type formulas.
		More precisely,
		for any positive integer $N$,
		each family of formulas depends on $2N+1$ parameters.
		The left-hand side of the formula is a sum over $N$-tuples of partitions,
		and the right-hand side of the formula is an infinite product.
	\end{abstract}

	\setcounter{section}{0}
	\setcounter{tocdepth}{2}
	

	\section{Introduction}
	
	The study of $q$-series originated with Euler,
	who established numerous identities relating infinite sums to infinite products.
	Many of his identities are closely related to the theory of symmetric functions.
	The so-called Nekrasov--Okounkov identity,
	which was derived by Nekrasov and Okounkov \cite{NO06} when they studied the Seiberg--Witten theory (see also the pioneer works by Westbury \cite{W06} and Han \cite{H08}),
	provides a sum formula for the powers of the Euler product:
	\begin{align}\label{eqn:NO in main}
		\sum_{\nu\in\mathcal{P}}
		s^{|\nu|}
		\cdot
		\prod_{(j,k)\in\nu}
		\frac{\big(h(j,k)+t\big)\big(h(j,k)-t\big)}
		{h(j,k)^2}
		=\prod\limits_{n\geq1} (1-s^n)^{t^2-1},
	\end{align}
	where $\mathcal{P}$ is the set of partitions and $h(j,k)$ is the hook-length.
	The left-hand side of the Nekrasov--Okounkov identity involves a sum over partitions and is weighted by a simple rational function of the hook-lengths.
	The right-hand side of this identity is a power of the Euler product,
	and at the same time, it is an infinite product.
	This identity is of significant importance,
	and various new proofs and generalizations have been explored in \cite{CO12,H08,H10,IKS,INRS,RW18,S10,W06} and the references therein.

	The topological vertex was introduced by Aganagic, Klemm, Mari\~{n}o and Vafa when they studied the A-model topological string amplitudes, as well as the open Gromov-Witten invariants, of toric Calabi--Yau threefolds (see \cite{AKMV,LLLZ,MOOP}).
	An important observation made by Okounkov, Reshetikhin and Vafa \cite{ORV} is that
	the topological vertex is connected to the enumeration of plane partitions,
	and subsequently to the so-called melting crystal model.
	As a result,
	the topological vertex possesses many symmetries and admits a formula via the vacuum expectation value.

	In this paper,
	by using the vacuum expectation value formula for the topological vertex and its rotation symmetry,
	we derive two families of Nekrasov--Okounkov type formulas.
	These formulas also establish identities between infinite sums over $N$-tuples of partitions and infinite product expressions,
	where $N$ is a positive integer.
	Each family of formulas depends on $2N+1$ parameters.
	The first family of Nekrasov--Okounkov type formulas is as follows.
	\begin{thm}\label{thm:main}
		For each positive integer $N$,
		and formal variables $q, Q_{1,i}, Q_{2,i}, i=1,...,N$,
		we have
		\begin{small}
			\begin{align}\label{eqn:main}
				\begin{split}
					&\sum_{\nu^1,...,\nu^N \in\mathcal{P}}
					\prod_{i=1}^N(-Q_{2,i})^{|\nu^i|}
					q^{-\|\nu^{i,t}\|^2}
					\prod_{(j,k)\in\nu^{i}}
					\frac{(1-Q_{1,i+1}q^{\nu^{i,t}_k+\nu^{i+1,t}_j-j-k+1})
						(1-Q_{1,i}q^{-\nu^{i-1}_k-\nu^{i}_j+j+k-1})}
					{(1-q^{-h(j,k)})^2}\\
					&\quad\quad\quad=\begin{cases}
						\frac{1}
						{\prod_{n\geq1}
							(1-\prod_{i=1}^NQ_{1,i}^nQ_{2,i}^n)}
						\cdot \prod_{1\leq i<k\leq N} f_{k,i,-1}
						\cdot\prod_{i,k=1}^N
						\prod_{j=0}^{\infty} f_{i,k,j}, \text{\ if\ }N\text{\ is\ even},\\
						\prod_{n\geq1}
						(1+\prod_{i=1}^NQ_{1,i}^{2n-1}Q_{2,i}^{2n-1})
						\cdot \prod_{1\leq i<k\leq N} g_{k,i,-1}
						\cdot\prod_{i,k=1}^N
						\prod_{j=0}^{\infty} g_{i,k,j}, \text{\ if\ }N\text{\ is\ odd},\\
					\end{cases}
				\end{split}
			\end{align}
		\end{small}
		where $\nu^t$ is the conjugation of $\nu$, $|\nu|=\sum_{i=1}^{l(\nu)}\nu_i, \|\nu\|^2=\sum_{i=1}^{l(\nu)}\nu_i^2$,
		and $Q_{1,N+1}=Q_{1,1}$, $\nu^{N+1}=\nu^1$, $\nu^0=\nu^N$,
		and $M(z;q)=\prod_{n=1}^\infty (1-zq^{-n})^n$,
		\begin{align}
			f_{i,k,j}=&\frac{M(Q_{1,i} Q_{1,k} a_{i,k,j};q)^{\epsilon_i \epsilon_k}
				\cdot M(a_{i,k,j};q)^{\epsilon_i \epsilon_k}}
			{M(Q_{1,k} a_{i,k,j};q)^{\epsilon_i \epsilon_k}
				\cdot M(Q_{1,i} a_{i,k,j};q)^{\epsilon_i \epsilon_k}}, \label{eqn:def f}\\
			g_{i,k,j}=&\frac{M\big((-1)^{j-1}Q_{1,i} Q_{1,k} a_{i,k,j};q\big)^{(-1)^{j-1}\epsilon_i \epsilon_k}
				\cdot M\big((-1)^{j-1}a_{i,k,j};q\big)^{(-1)^{j-1}\epsilon_i \epsilon_k}}
			{M\big((-1)^{j-1}Q_{1,k} a_{i,k,j};q\big)^{(-1)^{j-1}\epsilon_i \epsilon_k}
				\cdot M\big((-1)^{j-1}Q_{1,i} a_{i,k,j};q\big)^{(-1)^{j-1}\epsilon_i \epsilon_k}}, \label{eqn:def g}\\
			a_{i,k,j}=&\epsilon_i \epsilon_k\cdot
			Q_{2,k} \cdot \prod_{n=1}^{i-1}Q_{1,n}Q_{2,n}
			\cdot \prod_{n=k+1}^NQ_{1,n}Q_{2,n}
			\cdot \prod_{n=1}^N(Q_{1,n}Q_{2,n})^{j},
		\end{align}
		and $\epsilon_i=\delta_{i,odd}-\delta_{i,even}=1$ if $i$ is odd, otherwise it is $-1$.
		Both sides of the above equation \eqref{eqn:main} could be regarded as elements in the ring $\mathbb{C}[\![Q_{1,i},Q_{2,i},q^{-1}]\!]$.
	\end{thm}
	
	The result in Theorem \ref{thm:main} differs from those in \cite{IKS,INRS,NO06}, although all of them are derived by calculating certain partition functions.
	Our partition function considered in Theorem \ref{thm:main} (see Section \ref{sec:prove main} for more details) neither corresponds to a trace of certain operators in the Fock space,
	nor is obtained from the gluing rule in the theory of topological vertex \cite{AKMV} since the presence of certain partitions and their conjugations do not align with the gluing process.
	These distinctions become more evident in the following discussion.
	By taking a special limit (see subsection \ref{sec:app of main thm}), inspired by \cite{INRS,NO06},
	of the result in Theorem \ref{thm:main},
	we obtain the following proposition,
	\begin{prop}\label{cor:main}
		We have
		\begin{small}
			\begin{align}\label{eqn:main app in main}
				\begin{split}
					&\sum_{\nu^1,...,\nu^N \in\mathcal{P}}
					\prod_{i=1}^N (-s_i)^{|\nu^i|}
					\prod_{(j,k)\in\nu^{i}}
					\begin{small}
						\frac{\big(t_{i+1}+(\nu^{i,t}_k+\nu^{i+1,t}_j-j-k+1)\big)
							\big(t_i-(\nu^{i-1}_k+\nu^{i}_j-j-k+1)\big)
						}{h(j,k)^2}
					\end{small}\\
					&\quad=\begin{cases}
						\frac{1}
						{\prod\limits_{n\geq1}
							(1-s^n)}
						\cdot \prod\limits_{1\leq i<k\leq N} (1-b_{k,i,-1})^{\epsilon_i\epsilon_kt_it_k}
						\cdot\prod_{i,k=1}^N
						\prod_{j=0}^{\infty} (1-b_{i,k,j})^{\epsilon_i\epsilon_kt_it_k}, \text{\ if\ }N\text{\ is\ even},\\
						\prod\limits_{n\geq1}
						(1+s^{2n-1})
						\cdot \prod\limits_{1\leq i<k\leq N} (1-b_{k,i,-1})^{\epsilon_i\epsilon_kt_it_k}
						\cdot\prod\limits_{i,k=1}^N
						\prod\limits_{j=0}^{\infty} \big(1+(-1)^{j}b_{i,k,j}\big)^{\epsilon_j\epsilon_i\epsilon_kt_it_k}, \text{otherwise},
					\end{cases}
				\end{split}
			\end{align}
		\end{small}
		where $t_{N+1}=t_1, s=\prod_{i=1}^Ns_i$ and
		\begin{align*}
			b_{i,k,j}=\epsilon_i\epsilon_k
			\cdot\prod_{n=1}^{i-1}s_n
			\cdot\prod_{n=k}^N s_n
			\cdot\prod_{n=1}^N (s_n)^j.
		\end{align*}
	\end{prop}
	
	We hope that a direct combinatorial explanation exists for the above equation \eqref{eqn:main app in main},
	as its right-hand side consists solely of simple terms.
	The $N=1$ case of Proposition \ref{cor:main} above gives
	\begin{prop}
	We have
	\begin{align}\label{eqn:main app app in main}
		\sum_{\nu\in\mathcal{P}}
		s^{|\nu|}
		\prod_{(j,k)\in\nu}&
		\frac{\big(t+l(j,k)+l(k,j)+1\big)\big(t-a(j,k)-a(k,j)-1\big)}{h(j,k)^2}
		&
		=\prod\limits_{n\geq1}
		\frac{\big(1-s^{n}\big)^{(-1)^n t^2}}{(1+s^{n})},
	\end{align}
	where $a(j,k), l(j,k)$ represent the arm-length and leg-length,
	as explained in Subsection \ref{sec:schur}.
	\end{prop}
	Equation \eqref{eqn:main app app in main} above seems to be a conjugate version of the original Nekrasov--Okounkov formula.
	If one formally exchanges the positions of leg-length $l(k,j)$ and arm-length $a(k,j)$ in equation \eqref{eqn:main app app in main},
	then the left-hand side of it reduces to the original Nekrasov--Okounkov formula \eqref{eqn:NO in main}.
	However,
	to the best of the author's knowledge,
	this formula neither generalizes the Nekrasov--Okounkov formula nor serves as a corollary.
	This formula is expected to play a fundamental role similar to the original Nekrasov--Okounkov formula in various problems.
	Hence,
	it would be interesting to provide a direct combinatorial proof and refinement of equation \eqref{eqn:main app app in main} employing methods similar to those used in \cite{H10,S10,W06}.

	Our second family of Nekrasov--Okounkov type formulas provides a direct generalization of the results in \cite{INRS}.
	In that work,
	the authors proposed two methods to calculate the partition functions of the 5d U(1) theory and its generalizations.
	The first method yields an infinite sum formula,
	while the second provides an infinite product formula
	(some details can be seen in \cite{HIV} based on the Chern-Simons theory).
	In this paper,
	we give a $2N+1$ parameters' generalization of their formula using the vacuum expectation formula and rotation symmetry of the topological vertex.
	The cases $N=1,2$ were originally presented in \cite{INRS}.
	\begin{thm}\label{thm:main2}
		For each positive integer $N$,
		and formal variables $q, Q_{1,i}, Q_{2,i}, i=1,...,N$,
		we have
		\begin{small}
			\begin{align}\label{eqn:main2}
				\begin{split}
					\sum_{\nu^1,...,\nu^N \in\mathcal{P}}
					\prod_{i=1}^N (-Q_{2,i})^{|\nu^i|}
					q^{\kappa(\nu^i)/2-\|\nu^{i}\|^2}
					&\prod_{(j,k)\in\nu^i}
					\frac{(1-Q_{1,i+1}q^{\nu^{i}_j+\nu^{i+1,t}_k-j-k+1})
						(1-Q_{1,i}q^{-\nu^{i-1,t}_k-\nu^i_j+j+k-1})}{(1-q^{-h(j,k)})^2}\\
					&=\prod_{n\geq1}\frac{1}
					{ \big(1-\prod_{i=1}^NQ_{1,i}^nQ_{2,i}^n\big)}
					\cdot \prod_{1\leq i<k\leq N} \tilde{f}_{k,i,-1}
					\cdot\prod_{i,k=1}^N
					\prod_{j=0}^{\infty} \tilde{f}_{i,k,j},
				\end{split}
			\end{align}
		\end{small}
		where $Q_{1,N+1}=Q_{1,1}$, $\nu^{N+1}=\nu^1$, $\nu^0=\nu^N$, $M(z;q)=\prod_{n=1}^\infty (1-zq^{-n})^n$, and
		\begin{align*}
			\tilde{f}_{i,k,j}=&\frac{M(Q_{1,i}Q_{1,k}\tilde{a}_{i,k,j};q)
				\cdot M(\tilde{a}_{i,k,j};q)}
			{M(Q_{1,k} \tilde{a}_{i,k,j};q)
				\cdot M(Q_{1,i} \tilde{a}_{i,k,j};q)},\\
			\tilde{a}_{i,k,j}=&
			Q_{2,k} \cdot \prod_{n=1}^{i-1}Q_{1,n}Q_{2,n}
			\cdot \prod_{n=k+1}^NQ_{1,n}Q_{2,n}
			\cdot \prod_{n=1}^N(Q_{1,n}Q_{2,n})^{j}.
		\end{align*}
	\end{thm}
	
	The left-hand side of equation \eqref{eqn:main2} differs from that of equation \eqref{eqn:main}.
	The first difference lies in the powers of $q$:
	the left-hand side of equation \eqref{eqn:main} has the term $q^{-\|\nu^{i,t}\|^2}$,
	while the left-hand side of equation \eqref{eqn:main2} has the term $q^{\kappa(\nu^i)/2-\|\nu^{i}\|^2}$.
	The ratio of these two terms is $q^{\kappa(\nu^i)/2}\neq 1$,
	which is related to the so-called framing term in the theory of topological vertex \cite{AKMV}.
	The second difference involves the presence of $\nu^{i,t}$ and $\nu^{i}$.
	It is surprising that such simple differences in terms lead to vastly different infinite product formulas.
	It is worth noting that the infinite product formula in the theory of topological vertex is closely related to the integrality of the so-called Gopakumar-Vafa invariants (see \cite{HIV, LLZ}),
	and framing term naturally emerges in the gluing process (see \cite{AKMV,LLLZ}).
	
	Similarly to Proposition \ref{cor:main},
	certain limit of Theorem \ref{thm:main2} gives
	\begin{prop}\label{cor:main2}
		We have
		\begin{small}
			\begin{align*}
				&\sum_{\nu^1,...,\nu^N \in\mathcal{P}}
				\prod_{i=1}^N (-s_i)^{|\nu^i|}
				\prod_{(j,k)\in\nu^{i}}
				\frac{\big(t_{i+1}+(\nu^{i}_j+\nu^{i+1,t}_k-j-k+1)\big)
					\big(t_i-(\nu^{i-1,t}_k+\nu^i_j-j-k+1)\big)}
				{h(j,k)^2}\\
				&\quad\quad\quad\quad\quad\quad\quad\quad\quad\quad=
				\frac{1}
				{\prod\limits_{n\geq1}
					(1-s^n)}
				\cdot \prod\limits_{1\leq i<k\leq N} (1-\tilde{b}_{k,i,-1})^{t_it_k}
				\cdot\prod_{i,k=1}^N
				\prod_{j=0}^{\infty} (1-\tilde{b}_{i,k,j})^{t_it_k},
			\end{align*}
			where $t_{N+1}=t_1, s=\prod_{i=1}^Ns_i$ and
			\begin{align*}
				\tilde{b}_{i,k,j}=
				\prod_{n=1}^{i-1}s_n
				\cdot\prod_{n=k}^N s_n
				\cdot\prod_{n=1}^N (s_n)^j.
			\end{align*}
		\end{small}
	When $N=1$,
	this formula exactly matches with the Nekrasov--Okounkov formula \eqref{eqn:NO in main}.
	\end{prop}
	
	The rest of this paper is organized as follows.
	In Section \ref{sec:pre},
	we review the concepts of Schur functions, vertex operators and the topological vertex.
	These concepts are fundamental in studying the generalized Nekrasov--Okounkov type formulas presented in this paper.
	Section \ref{sec:prove main} and Section \ref{sec:prove main2} are dedicated to proving Theorem \ref{thm:main} and Theorem \ref{thm:main2} respectively.
	These theorems establish connections between two families of infinite sums over tuples of partitions and two families of infinite product formulas.
	These two sections also include the applications of these two families of formulas,
	which prove Proposition \ref{cor:main} and Proposition \ref{cor:main2}

	\section{Preliminaries}
	\label{sec:pre}
	In this section,
	we first review the definitions and basic properties of Schur functions, vertex operators and the topological vertex.
	
	\subsection{Integer partitions and Schur functions}
	\label{sec:schur}
	In this subsection,
	we review the definition of Schur functions and skew Schur functions.
	We recommend Macdonald's book \cite{Mac}.
	
	A partition is a sequence of weakly decreasing positive integers $\lambda=(\lambda_1,...,\lambda_l)$.
	Its size is $|\lambda|=\sum_{i=1}^l \lambda_i$ and its length is $l$.
	We also set $\lambda_i=0$ for all $i>l$.
	For each partition $\lambda$,
	there is a corresponding Young diagram,
	which has $\lambda_i$ boxes in its $i-$th row.
	For example,
	the following Figure \ref{eqn:Yd 5431} is the Young diagram corresponding to the partition $(5,4,3,1)$.
	\begin{figure}[htbp]
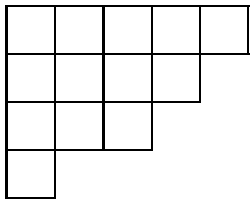

		\ydiagram{5,4,3,1}
		\caption{Young diagram corresponding to $(5,4,3,1)$}
		\label{eqn:Yd 5431}
	\end{figure}
	
	The partition $\lambda^t$ is the conjugate of $\lambda$ and is defined by
	\begin{align*}
		\lambda^t_i=\#\{j|\lambda_j\geq i\}.
	\end{align*}
	The Young diagram of $\lambda^t$ is the transpose of the Young diagram corresponding to $\lambda$.
	For example,
	the conjugation of the partition $(5,4,3,1)$ in Figure \ref{eqn:Yd 5431} is $(4,3,3,2,1)$.
	
	For each box $(i,j)$ in the Young diagram of partition $\lambda$,
	denote the content by $c(i,j)=j-i$ and the hook-length by $h(i,j)=\lambda_i+\lambda^t_j-i-j+1$.
	It is also useful to introduce the arm-length $a(i,j)=\lambda_i-j$ and the leg-length $l(i,j)=\lambda^t_j-i$ (see Chapter VI.6 (6.14) in \cite{Mac}).
	Then $h(i,j)=a(i,j)+l(i,j)+1$.
	Notice that,
	the content, hook-length, arm-length and leg-length are well-defined even when $(i,j)$ is not a box in the Young diagram of $\lambda$ by setting $\lambda_i=0$ for all $i>l(\lambda)$.
	
	The Schur function associated with a partition $\lambda=(\lambda_1,...,\lambda_l)$ is defined as follows
	\begin{align*}
		s_\lambda
		=s_\lambda(x_1,x_2,...)
		=\lim_{n\rightarrow\infty}
		\frac{\det(x_i^{\lambda_j+n-j})_{1\leq i,j\leq n}}
		{\det(x_i^{n-j})_{1\leq i,j\leq n}}.
	\end{align*}
	The above limit is well-defined in a suitable ring of symmetric functions in countably many variables (see Chapter I.2 in \cite{Mac}).
	Thus, Schur functions should be considered as formal symmetric functions with respect to the variables $\{x_i\}_{1\leq i<\infty}$.
	On the other hand,
	by using the power-sum coordinates
	\begin{align*}
		p_k=p_k(x_1,x_2,...)=\sum_{i=1}^\infty x_i^k,
		\quad 1\leq k<\infty,
	\end{align*}
	the Schur function $s_\lambda$ could also be regarded as a homogeneous polynomial of degree $|\lambda|=\sum_{i=1}^{l(\lambda)}\lambda_i$
	in the polynomial ring $\mathbb{C}[p_1,p_2,...]$
	where $\deg p_k=k$ (or $\deg x_i=1$).
	Denote by $\mathcal{P}=\{\lambda|\lambda=(\lambda_1,...,\lambda_l)\}$ the set of partitions.
	Then it is well known that $\{s_\lambda\}|_{\lambda\in\mathcal{P}}$ is a basis of $\mathbb{C}[p_1,p_2,...]$.

	The skew Schur function is a generalization of the Schur function, labeled by two partitions.
	For partitions $\lambda, \mu\in\mathcal{P}$,
	the corresponding skew Schur function is defined as
	\begin{align*}
		s_{\lambda/\mu}=\sum_{\nu} c^{\lambda}_{\mu \nu} s_{\nu},
	\end{align*}
	where $\{c^{\lambda}_{\mu \nu}\}$ are the Littlewood-Richardson coefficients defined by $s_{\mu} s_{\nu}=\sum_{\lambda} c^{\lambda}_{\mu \nu} s_{\lambda}$.
	Since $c^\lambda_{\mu \nu}=0$ unless $|\lambda|=|\mu|+|\nu|$,
	$s_{\lambda/\mu}$ is a homogeneous polynomial of degree $|\lambda|-|\mu|$.

	In this paper,
	the evaluations of Schur functions are always with respect to the $x_i$ variables unless otherwise specified.
	That is,
	\[s_\lambda(a_1,a_2,...)
	=s_\lambda|_{x_i\rightarrow a_i}\]
	denotes the evaluation of $s_\lambda$ at $x_i=a_i$,
	where $(a_1,a_2,...)$ is a sequence of numbers or formal variables.
	In general, the evaluations of Schur functions may not be well-defined since $s_\lambda$ is a formal function.
	However, the special evaluations used in the theory of topological vertex are always meaningful \cite{AKMV}.
	A useful case is the following.
	Let $\rho=(-1/2,-3/2,...)$
	and $q^{\rho}=(q^{-1/2},q^{-3/2},...)$,
	then by I.3 Example 1 in \cite{Mac},
	we have
	\begin{align}\label{eqn:schur q^rho}
		s_{\lambda}(q^{\rho})
		=s_{\lambda}(q^{-1/2},q^{-3/2},...)
		=q^{-n(\lambda)-|\lambda|/2}\prod_{(i,j)\in\lambda} \frac{1}{1-q^{-h(i,j)}},
	\end{align}
	where $n(\lambda)=\sum_{i=1}^{l(\lambda)} (i-1)\lambda_i$.
	Both sides of equation \eqref{eqn:schur q^rho} can be regarded as a rational function of $q$ or a formal power series in the ring $\mathbb{C}[\![q^{-1}]\!]$.
	In this case $x_i=q^{-i+1/2}$,
	the power-sum coordinates are given by
	\begin{align*}
		p_n(x_i)=\frac{q^{-n/2}}{1-q^{-n}}
		=\frac{1}{q^{n/2}-q^{-n/2}}.
	\end{align*}
	
	Note that,
	let $\kappa(\lambda)=\sum_{i=1}^{l(\lambda)} \lambda_i(\lambda_i-2i+1),
	\binom{\lambda}{2}=\sum_{i=1}^{l(\lambda)} \binom{\lambda_i}{2}$ and
	$\|\lambda\|^2=\sum_{i=1}^{l(\lambda)}\lambda_i^2$,
	then
	\begin{align*}
		n(\lambda)
		=\sum_{i=1}^{l(\lambda)}\sum_{j=1}^{\lambda_i} (i-1)
		=\sum_{j=1}^{l(\lambda^t)}\sum_{i=1}^{\lambda^t_j} (i-1)
		=\binom{\lambda^t}{2}
		=\|\lambda^t\|^2/2-|\lambda|/2,
	\end{align*}
	and
	\begin{align*}
		\frac{\kappa(\lambda)}{2}
		=\frac{1}{2}\sum_{i=1}^{l(\lambda)}\lambda_i^2
		-\sum_{i=1}^{l(\lambda)}i\lambda_i
		+|\lambda|/2
		=\frac{\|\lambda\|^2}{2}
		-\frac{\|\lambda^t\|^2}{2}
		=\binom{\lambda}{2}-\binom{\lambda^t}{2}.
	\end{align*}
	
	\subsection{Vertex operators}
	In this subsection,
	we review the concept of free charged fermions.
	The vertex operators are certain generating functions of these fermions. 
	
	\subsubsection{Free charged fermions}
	The free charged fermions are operators $\{\psi_k,\psi^*_k\}_{k\in\mathbb{Z}+\half}$ that satisfy the following commutation relations
	\begin{align}\label{eqn:comm psipsi*}
		[\psi_{r},\psi_{s}]_+=[\psi^*_{r},\psi^*_{s}]_+=0,
		\quad [\psi_{r},\psi^*_{s}]_+=\delta_{r+s,0}\cdot \text{Id},
	\end{align}
	where the bracket $[A,B]_+$ is defined as $AB+BA$
	and $\text{Id}$ is the identity operator.
	There is a canonical representation for the algebra generated by all fermions and the identity operator (see, for example, \cite{DJM}).
	The representation space $\mathcal{F}$ is called the fermionic Fock space.
	It is spanned by the vacuum vector $|0\rangle$ that satisfies
	\begin{align}\label{eqn:annihilators}
		\psi_k |0\rangle=0,
		\quad\quad \psi^*_k |0\rangle=0,
		\quad\quad \forall\ k>0.
	\end{align}
	That is,
	\begin{align*}
		\mathcal{F}=
		\text{span}_{\mathbb{C}}\{\psi_{k_1}\cdots\psi_{k_n} 
		\psi^*_{k'_1}\cdots\psi^*_{k'_m} |0\rangle\ 
		|\ k_1<\cdots<k_n<0, k'_1<\cdots<k'_m<0\},
	\end{align*}
	and the fermions act on $\mathcal{F}$ in the canonical way, according to the commutation relations \eqref{eqn:comm psipsi*}.
	
	We assume that $\psi^*_{-k}$ is the dual operator of $\psi_k$, and denote $\mathcal{F}^*$ as the dual space of the fermionic Fock space.
	As a result,
	$\mathcal{F}^*$ is spanned by the actions of fermions on the dual vacuum vector $\langle0|$,
	where $\langle0|$ is the dual of $|0\rangle$.
	The vacuum expectation value is then defined in terms of the action of $\mathcal{F}^*$ on $\mathcal{F}$ and is denoted by
	\begin{align*}
		\langle0|
		\tilde{\psi}_{j_1}\cdots\tilde{\psi}_{j_n} 
		|0\rangle\in\mathbb{C},
	\end{align*}
	where $\tilde{\psi_r}$ can be either $\psi_r$ or $\psi^*_r$,
	and $j_k\in\mathbb{Z}+\half$.
	The vacuum expectation value can be determined by the commutation relations \eqref{eqn:comm psipsi*}, equation \eqref{eqn:annihilators}, linearity and the initial value $\langle0|0\rangle=1$.
	
	The normal ordering of product of fermions is then defined as follows,
	\begin{align*}
		:\tilde{\psi}_r\tilde{\psi}_s:=\tilde{\psi}_r\tilde{\psi}_s-\langle0|\tilde{\psi}_r\tilde{\psi}_s|0\rangle\cdot \text{Id},
	\end{align*}
	where $\tilde{\psi}_r$ can be either $\psi_r$ or $\psi^*_r$.

	\subsubsection{The vertex operators}
	The Heisenberg operators are defined by
	\begin{align*}
		\alpha_n = \sum_{k\in\mathbb{Z}+\half}
		:\psi_{-k}\psi^*_{k+n}:,
		\quad n\in\mathbb{Z}.
	\end{align*}
	Operators $\alpha_n$ and $\alpha_{-n}$ are dual to each other, and they satisfy the following standard commutation relation
	\begin{align*}
		[\alpha_m,\alpha_n]=m\delta_{m+n,0}\cdot\text{Id}.
	\end{align*}
	The so-called vertex operators (see, for example, \cite{DJM}) are certain exponential generating functions of those Heisenberg operators defined by
	\begin{align*}
		\Gamma_\pm(z)
		=\exp\Big(\sum_{n=1}^\infty \frac{z^n}{n} \alpha_{\pm n}\Big).
	\end{align*}
	They are useful in boson-fermionic correspondence and can be used to generate Schur functions and skew Schur functions.
	Another important differential operator is the so-called energy operator, defined by
	\begin{align*}
		L_0=\sum_{k\in\mathbb{Z}+\half}
		k:\psi_{-k}\psi^*_{k}:.
	\end{align*}
	The eigenvector subspaces of this energy operator give a decomposition of the fermionic Fock space $\mathcal{F}$ in terms of the energy,
	which is equivalent to the degree under boson-fermionic correspondence.
	
	The commutation relations of vertex operators and energy operator are given by (see, for example, \cite{O01})
	\begin{align}\label{eqn:comm g+-}
		\Gamma_+(z)\Gamma_-(w)=
		\frac{1}{1-zw}\Gamma_-(w)\Gamma_+(z)
	\end{align}
	and
	\begin{align}\label{eqn:comm gL0}
		q^{L_0}\Gamma_+(z)=\Gamma_+(q^{-1}z)q^{L_0},
		\quad\quad
		q^{L_0}\Gamma_-(z)=\Gamma_-(qz)q^{L_0}.
	\end{align}

	\subsection{Schur functions in terms of fermions}
	The Schur functions can be realized in terms of the vacuum expectation values of fermions.
	To be precise,
	for each partition $\lambda=(\lambda_1,...,\lambda_l)$,
	denote by $(m_1,...,m_r|n_1,...,n_r)$ the Frobenius notation of $\lambda$.
	The $r$ is determined by the maximal integer satisfying $\lambda_r-r\geq0, \lambda^t_r-r\geq0$
	and the $m_i$ and $n_i$ are
	\begin{align*}
		m_i=\lambda_i-i,
		\quad n_i=\lambda^t_i-i.
	\end{align*}
	Then the vector $|\lambda\rangle\in\mathcal{F}$ corresponding to the Schur function $s_\lambda$ is defined by
	\begin{align*}
		|\lambda\rangle
		=(-1)^{n_1+\dots+n_r}
		\cdot\psi_{-m_1-\half}\psi^*_{-n_1-\half}
		\cdots
		\psi_{-m_r-\half}\psi^*_{-n_r-\half}
		|0\rangle.
	\end{align*}
	Denote by $\langle\lambda|$ the dual vector of $|\lambda\rangle$.
	For the sequence of symmetric variables $\bm x=(x_1,x_2,...)$,
	the corresponding vertex operators are
	\begin{align*}
		\Gamma_\pm(\bm x)
		=\prod_{i=1}^\infty \Gamma_\pm(x_i).
	\end{align*}
	Then, the famous boson-fermionic correspondence gives
	\begin{align*}
		s_\lambda(\bm x)
		=\langle0|\Gamma_+(\bm x)|\lambda\rangle
		=\langle\lambda|\Gamma_-(\bm x)|0\rangle.
	\end{align*}
	In general,
	the actions of vertex operators on the vectors corresponding to Schur functions are given by the following (see, for example, equation (A.15) in \cite{O01})
	\begin{align}\label{eqn:gamma- action}
		\Gamma_-(\bm x)|\lambda\rangle
		=\sum_{\mu}s_{\mu/\lambda}(\bm x)|\mu\rangle,
		\quad\quad
		\langle\lambda|\Gamma_+(\bm x)
		=\sum_{\mu}s_{\mu/\lambda}(\bm x)\langle\mu|
	\end{align}
	and
	\begin{align}\label{eqn:gamma+ action}
		\Gamma_-(\bm x)\langle\lambda|
		=\sum_{\mu}s_{\lambda/\mu}(\bm x)\langle\mu|,
		\quad\quad\Gamma_+(\bm x)|\lambda\rangle
		=\sum_{\mu}s_{\lambda/\mu}(\bm x)|\mu\rangle.
	\end{align}
	The above equations can be sufficiently used to generate skew Schur functions,
	which are used in defining the topological vertex \cite{AKMV}.
	The precise relation between such vertex operators and the topological vertex was studied in \cite{ORV} and will be reviewed in Subsection \ref{subsec:tv}.

	\subsection{The $w$ transform}
	\label{sec:w trans}
	The $w$ transform is an involution on the ring of symmetric functions in variables $\{x_i\}_{i=1}^\infty$,
	or equivalently, the ring of polynomials $\mathbb{C}[p_1,p_2,...]$.
	It is very closely related to taking conjugation of partitions and provides some symmetries of the topological vertex, as we will see in Section \ref{sec:prove main} and Section \ref{sec:prove main2}.
	
	First,
	the $w$ transform is defined as a homomorphism of the ring $\mathbb{C}[p_1,p_2,...]$ 
	satisfying (see I.2 in \cite{Mac})
	\[w(s_{(a)})=s_{(a)^t},\]
	where $s_{(a)}$ is the so called complete symmetric function and $s_{(a)^t}$ is the elementary symmetric function in literature.
	An important fact is that the above formula still holds for the general Schur functions and even the skew Schur functions.
	That is to say (see equation (5.6) in Chapter I in \cite{Mac})
	\begin{align}
		w(s_{\lambda/\mu})
		=s_{\lambda^t/\mu^t}.
	\end{align}
	As a consequence,
	the involution $w$ should be understood as taking conjugation of partitions when meeting the skew Schur functions.
	
	On the other hand,
	under the power sum coordinates $p_k$,
	the involution $w$ is given by (equation (2.13) in Chapter I in \cite{Mac})
	\begin{align}
		w(p_n)=(-1)^{n-1}p_n,
		\quad\text{and}\quad
		w(p_{n_1}\cdots p_{n_k})=(-1)^{n_1+\cdots n_k-k}
		p_{n_1}\cdots p_{n_k}.
	\end{align}
	The result is that,
	the influence of the $w$ transform on the vertex operators is
	\begin{align*}
		w:  \Gamma_\pm(x_1,x_2,...)
		\mapsto
		\Gamma_\pm(-x_1,-x_2,...)^{-1}
	\end{align*}
	since $(\Gamma_\pm)_{p_n\mapsto-p_n}=(\Gamma_\pm)^{-1}$
	and $p_n(-x_1,-x_2,...)=(-1)^n p_n(x_1,x_2,...)$.
	Then combing with equations \eqref{eqn:gamma- action} and \eqref{eqn:gamma+ action},
	we can obtain a series of identities involving vacuum expectation values of vertex operators.
	For example,
	\begin{align}\label{eqn:mu,mu^t -1}
		\langle\mu|
		\Gamma_-(\bm x)^{\epsilon_i}
		\Gamma_+(\bm y)^{\epsilon'_i}
		|\lambda\rangle
		=\langle\mu^t|
		\Gamma_-(-\bm x)^{-\epsilon_i}
		\Gamma_+(-\bm y)^{-\epsilon'_i}
		|\lambda^t\rangle
	\end{align}
	where $\epsilon_i$ and $\epsilon'_i$ could be arbitrary $\pm1$.

	\subsection{Topological vertex}
	\label{subsec:tv}
	The topological vertex was proposed by Aganagic--Klemm--Mari\~{n}o--Vafa \cite{AKMV} for studying the A-model topological string amplitudes of smooth toric Calabi--Yau threefolds,
	as well as the open Gromov--Witten invariants of corresponding manifolds (see \cite{LLLZ} and \cite{MOOP} for a mathematical theory).
	
	In this paper,
	we directly use the combinatorial properties of the topological vertex and ignore its algebraic geometric meaning.
	We use the following expression as the definition of the topological vertex
	\begin{align}\label{eqn:def C}
		C_{\lambda,\mu,\nu}(q)
		= q^{\kappa(\lambda)/2+\kappa(\nu)/2}
		\cdot s_{\nu^t} (q^{\rho})
		\cdot \sum_\eta s_{\lambda^t /\eta} (q^{\nu + \rho})
		s_{\mu /\eta} (q^{\nu^t + \rho}),
	\end{align}
	where $\rho = (-\half, -\frac{3}{2}, -\frac{5}{2}, \cdots)$.
	Our definition \eqref{eqn:def C} of the topological vertex follows from \cite{ORV},
	(which is slightly different from the notation in \cite{INRS}).
	
	In terms of the transfer matrix method to the crystal melting model,
	Okounkov--Reshetikhin--Vafa showed that the partition function of certain plane partitions $P(\lambda,\mu,\nu)$ can be calculated by a formula in terms of vacuum expectation value,
	and its connection to the topological vertex is (see equation (3.21) in \cite{ORV})
	\begin{align*}
		C_{\lambda,\mu,\nu}(q^{-1})
		=q^{\|\lambda^t\|^2/2+\|\mu^t\|^2/2+\|\nu^t\|^2/2}
		\cdot \prod_{n>0}(1-q^n)^n
		\cdot P(\lambda,\mu,\nu).
	\end{align*}
	
	A special case of the vacuum expectation value formula (equation (3.17) in \cite{ORV}) is
	\begin{align}\label{eqn:clm0=<>}
		C_{\lambda,\mu,\emptyset}(q)
		=&q^{\kappa(\lambda)/2}
		\prod_{n>0}(1-q^{-n})^n
		\cdot\langle\lambda^t|
		\Gamma_+(q^\rho)
		\Gamma_-(q^\rho)|\mu\rangle\\
		=&q^{\kappa(\lambda)/2}
		\langle\lambda^t|
		\Gamma_-(q^\rho)
		\Gamma_+(q^\rho)|\mu\rangle,
	\end{align}
	where we have used the commutation relation \eqref{eqn:comm g+-}.
	
	The topological vertex $C_{\lambda,\mu,\nu}(q)$ possess many symmetries.
	First,
	by the rotation symmetry of plane partitions,
	we have
	\begin{align}\label{eqn:rotation symm C}
		C_{\lambda,\mu,\nu}(q)
		=C_{\mu,\nu,\lambda}(q)
		=C_{\nu,\lambda,\mu}(q).
	\end{align}
	By taking mirror of plane partitions,
	we have $P(\lambda,\mu,\nu)=P(\mu^t,\lambda^t,\nu^t)$,
	which gives
	\begin{align*}
		q^{\|\lambda^t\|^2/2+\|\mu^t\|^2/2+\|\nu^t\|^2/2}
		C_{\lambda,\mu,\nu}(q)
		=q^{\|\lambda\|^2/2+\|\mu\|^2/2+\|\nu\|^2/2}
		C_{\mu^t,\lambda^t,\nu^t}(q)
	\end{align*}
	and thus
	\begin{align*}
		C_{\lambda,\mu,\nu}(q)
		=q^{\kappa(\lambda)/2+\kappa(\mu)/2+\kappa(\nu)/2}
		C_{\mu^t,\lambda^t,\nu^t}(q).
	\end{align*}

	\section{Proof of the first family of Nekrasov--Okounkov type formulas}\label{sec:prove main}
	In this section,
	we prove the Theorem \ref{thm:main} and Proposition \ref{cor:main}.
	
	For each positive integer $N$,
	$q$ and $Q_{1,i}, Q_{2,i}, 1\leq i \leq N$ are formal variables.
	We consider the following partition function,
	\begin{align}\label{eqn:def Z_N}
		Z_N(Q_{1,i},Q_{2,i};q)
		=\sum_{\mu^1,\nu^1,...,\mu^N,\nu^N \in\mathcal{P}}
		\prod_{i=1}^N(-Q_{1,i})^{|\mu^i|}(-Q_{2,i})^{|\nu^i|}
		C_{\emptyset,\mu^{i,t},\nu^i}(q) C_{\emptyset,\mu^{i+1},\nu^{i}}(q),
	\end{align}
	where we use the notation $\mu^{N+1}=\mu^1$.
	$Z_N$ should be regarded as a formal power series in the ring $\mathbb{C}[\![Q_{1,i},Q_{2,i},q^{-1}]\!]$.
	We will use two different methods to calculate the above partition function below.
	
	\begin{rmk}
		It will be interesting to ask whether the above partition function $Z_N$ can be explained as a closed string amplitude of a mathematical physical model regarding the gluing rule of topological vertex \cite{AKMV}.
		The difficulty here is that,
		the twice appearances of $\nu^i$ are $\nu^i$ itself,
		but for $\mu^{i}$,
		the first appearance is $\mu^{i,t}$ and the second appearance is $\mu^{i}$,
		which aligns with the gluing rule.
	\end{rmk}
	
	\subsection{Infinite product formula for $Z_N$}
	
	The following lemma helps deal with even $N$ case of Theorem \ref{thm:main} and in Section \ref{sec:prove main2}.
	\begin{lem}\label{lem:mu,g-Qg+ ep,mu}
		We assume $|Q|<1$, then
		\begin{align}\label{eqn:mu,g-Qg+ ep,mu}
			&\sum_{\mu\in\mathcal{P}}
			\langle \mu|
			\prod_{i=1}^L
			\Gamma_-(x_{1,i}\cdot q^{\rho})^{\epsilon_{1,i}}
			\cdot Q^{{L_0}} \cdot
			\prod_{i=1}^L
			\Gamma_+(y_{1,i}\cdot q^{\rho})^{\epsilon_{2,i}}
			|\mu\rangle\\
			&\quad\quad\quad\quad\quad\quad\quad\quad\quad\quad\quad
			=\prod_{n\geq1}\frac{1}{(1-Q^n)}
			\cdot \prod_{i,k=1}^L
			\prod_{j=0}^{\infty}
			\frac{1}
			{M(Q^jx_{1,i}y_{1,k};q)^{\epsilon_{1,i}\epsilon_{2,k}}}
		\end{align}
		as elements in the ring $\mathbb{C}[\![Q,x_{1,i},y_{1,i},q^{-1}]\!]$,
		where $\epsilon_{k,i}$ are arbitrary $\pm1$, and $M(z;q)=\prod_{n=1}^\infty (1-zq^{-n})^n$.
	\end{lem}
	{\bf Proof:}
	First,
	since $\{s_{\lambda}\}_{\lambda\in\mathcal{P}}$ is a basis of the polynomial ring $\mathbb{C}[p_1,p_2,...]$,
	the following operator
	\[\sum_{\mu'\in\mathcal{P}}|\mu'\rangle\langle\mu'|\]
	could be regarded as an identity operator on the ring $\mathbb{C}[p_1,p_2,...]$.
	We insert it in front of $Q^{L_0}$ in equation \eqref{eqn:mu,g-Qg+ ep,mu},
	and taking the summation over $\mu\in\mathcal{P}$.
	Then equation \eqref{eqn:mu,g-Qg+ ep,mu} is equal to
	\begin{align*}
		&\sum_{\mu,\mu'\in\mathcal{P}}
		\langle \mu|
		\prod_{i=1}^L
		\Gamma_-(x_{1,i}\cdot q^{\rho})^{\epsilon_{1,i}}
		|\mu'\rangle \cdot \langle\mu'|
		Q^{{L_0}} \cdot
		\prod_{i=1}^L 
		\Gamma_+(y_{1,i}\cdot q^{\rho})^{\epsilon_{2,i}}
		|\mu\rangle\\
		&\quad\quad\quad\quad\quad\quad\quad
		=\sum_{\mu'\in\mathcal{P}}
		\langle\mu'|
		Q^{{L_0}} \cdot
		\prod_{i=1}^L 
		\Gamma_+(y_{1,i}\cdot q^{\rho})^{\epsilon_{2,i}}
		\cdot
		\prod_{i=1}^L
		\Gamma_-(x_{1,i}\cdot q^{\rho})^{\epsilon_{1,i}}
		|\mu'\rangle.
	\end{align*}
	Using commutation relations \eqref{eqn:comm g+-} and \eqref{eqn:comm gL0},
	we change the orders of $\Gamma_\pm$ and $Q^{L_0}$.
	Then equation \eqref{eqn:mu,g-Qg+ ep,mu} reduces to
	\begin{align*}
		&\prod_{i,k=1}^L
		\frac{1}
		{M(x_{1,i}y_{1,k};q)^{\epsilon_{1,i}\epsilon_{2,k}}}
		\cdot \sum_{\mu'\in\mathcal{P}}
		\langle\mu'|
		\prod_{i=1}^L
		\Gamma_-(Qx_{1,i}\cdot q^{\rho})^{\epsilon_{1,i}}
		\cdot Q^{{L_0}} \cdot
		\prod_{i=1}^L
		\Gamma_+(y_{1,i}\cdot q^{\rho})^{\epsilon_{2,i}}
		\cdot
		|\mu'\rangle.
	\end{align*}
	Next, we use the same method to insert the identity operator 
	$\sum_{\mu\in\mathcal{P}}|\mu\rangle\langle\mu|$
	behind $Q^{L_0}$ in the above equation.
	Then, one can obtain that
	the equation \eqref{eqn:mu,g-Qg+ ep,mu} is equal to
	\begin{align*}
		&\prod_{i,k=1}^L
		\prod_{j=0}^1
		\frac{1}
		{M(Q^jx_{1,i}y_{1,k};q)^{\epsilon_{1,i}\epsilon_{2,k}}}\\
		&\quad\quad\quad\quad\quad\cdot \sum_{\mu\in\mathcal{P}}
		\langle\mu|
		\prod_{i=1}^L
		\Gamma_-(Qx_{1,i}\cdot q^{\rho})^{\epsilon_{1,i}}
		\cdot Q^{{L_0}} \cdot
		\prod_{i=1}^L 
		\Gamma_+(Qy_{1,i}\cdot q^{\rho})^{\epsilon_{2,i}}
		\cdot
		|\mu\rangle.
	\end{align*}
	One can notice that,
	after inserting the identity operator twice,
	the operators $\Gamma_\pm(\cdot)$ will become $\Gamma_\pm(Q\cdot)$.
	Thus,
	for any positive integer $K$,
	one can repeat the above process $K$ times.
	The result is that,
	equation \eqref{eqn:mu,g-Qg+ ep,mu} is equal to
	\begin{align*}
		&\prod_{i,k=1}^\infty
		\prod_{j=0}^{2K-1}
		\frac{1}
		{M(Q^jx_{1,i}y_{1,k};q)^{\epsilon_{1,i}\epsilon_{2,k}}}\\
		&\quad\quad\quad\quad\quad\cdot \sum_{\mu\in\mathcal{P}}
		\langle\mu|
		\prod_{i=1}^L
		\Gamma_-(Q^Kx_{1,i}\cdot q^{\rho})^{\epsilon_{1,i}}
		\cdot Q^{{L_0}} \cdot
		\prod_{i=1}^L 
		\Gamma_+(Q^Ky_{1,i}\cdot q^{\rho})^{\epsilon_{2,i}}
		|\mu\rangle.
	\end{align*}
	By letting $K\rightarrow+\infty$,
	because of $|Q|<1$,
	\begin{align*}
		\Gamma_-(Q^Kx_{1,i}\cdot q^{\rho})^{\epsilon_{1,i}}
		=\Gamma_+(Q^Ky_{1,i}\cdot q^{\rho})^{\epsilon_{2,i}}
		\rightarrow\text{Id}
	\end{align*}
	the identity operator.
	Then,
	this lemma is proved with the following identity:
	\begin{align*}
		\sum_{\mu\in\mathcal{P}}
		\langle\mu|
		Q^{{L_0}}
		|\mu\rangle
		=\sum_{\mu\in\mathcal{P}} Q^{|\mu|}
		=\prod_{n\geq1}\frac{1}{(1-Q^n)}.
	\end{align*}
	$\Box$
	
	A special case of the above Lemma is the following,
	\begin{cor}\label{cor:mu,g--Qg++ ep,mu}
		We assume $|Q|<1$, then
		\begin{align}\label{eqn:mu,g--Qg++ ep,mu}
			&\sum_{\mu\in\mathcal{P}}
			\langle \mu|
			\prod_{i=1}^N
			\big(\Gamma_-(x_{1,i}\cdot q^{\rho})^{\epsilon_{1,i}}
			\Gamma_-(x_{2,i}\cdot q^{\rho})^{\epsilon_{2,i}}\big)
			\cdot Q^{{L_0}} \cdot
			\prod_{i=1}^N 
			\big(\Gamma_+(y_{1,i}\cdot q^{\rho})^{\epsilon_{3,i}}
			\Gamma_+(y_{2,i}\cdot q^{\rho})^{\epsilon_{4,i}}\big)
			|\mu\rangle \nonumber\\
			&\quad\quad
			=\prod_{n\geq1}\frac{1}{(1-Q^n)}
			\cdot \prod_{i,k=1}^N
			\prod_{j=0}^{\infty}
			\frac{M(Q^jx_{1,i}y_{2,k};q)^{-\epsilon_{1,i}\epsilon_{4,k}}
				\cdot M(Q^jx_{2,i}y_{1,k};q)^{-\epsilon_{2,i}\epsilon_{3,k}}}
			{M(Q^jx_{1,i}y_{1,k};q)^{\epsilon_{1,i}\epsilon_{3,k}}
				\cdot M(Q^jx_{2,i}y_{2,k};q)^{\epsilon_{2,i}\epsilon_{4,k}}}.
		\end{align}
	\end{cor}
	{\bf Proof:}
	Just by applying the $L=2N$ case of Lemma \ref{lem:mu,g-Qg+ ep,mu}.
	$\Box$

	The following lemma is useful for the odd $N$ case of Theorem \ref{thm:main}.
	\begin{lem}\label{lem:mu,g-Qg+ ep,mu^t}
		We assume $|Q|<1$, then
		\begin{align}\label{eqn:mu,g-Qg+ ep,mu^t}
			&\sum_{\mu\in\mathcal{P}}
			\langle \mu|
			\prod_{i=1}^L
			\Gamma_-(x_{1,i}\cdot q^{\rho})^{\epsilon_{1,i}}
			\cdot Q^{{L_0}} \cdot
			\prod_{i=1}^L
			\Gamma_+(y_{1,i}\cdot q^{\rho})^{\epsilon_{2,i}}
			|\mu^t\rangle\\
			&\quad\quad\quad\quad\quad\quad
			=\prod_{n\geq1}(1+Q^{2n-1})
			\cdot \prod_{i,k=1}^L
			\prod_{j=0}^{\infty}
			\frac{1}
			{M\big((-1)^{j-1}Q^jx_{1,i}y_{1,k};q\big)^{(-1)^{j-1}\epsilon_{1,i}\epsilon_{2,k}}}.
		\end{align}
	\end{lem}
	{\bf Proof:}
	The method of proving this lemma is similar to that used in proving Lemma \ref{lem:mu,g-Qg+ ep,mu},
	so we explain the key steps below.
	First,
	by inserting the identity operator
	$\sum_{\mu'\in\mathcal{P}}|\mu'\rangle\langle\mu'|$
	in front of $Q^{L_0}$ in equation \eqref{eqn:mu,g-Qg+ ep,mu^t},
	it is equal to
	\begin{align}\label{eqn:main lem step 1}
		\sum_{\mu,\mu' \in\mathcal{P}}
		\langle \mu|
		\prod_{i=1}^L
		\Gamma_-(x_{1,i}\cdot q^{\rho})^{\epsilon_{1,i}}
		|\mu'\rangle
		\cdot \langle\mu'|
		Q^{{L_0}} \cdot
		\prod_{i=1}^L
		\Gamma_+(y_{1,i}\cdot q^{\rho})^{\epsilon_{2,i}}
		|\mu^t\rangle.
	\end{align}
	For the first factor in the above equation,
	we use the $w$-transform to rewrite it as
	(see equation \eqref{eqn:mu,mu^t -1})
	\begin{align*}
		\langle \mu^t|
		\prod_{i=1}^L
		\Gamma_-(-x_{1,i}\cdot q^{\rho})^{-\epsilon_{1,i}}
		|\mu^{',t}\rangle.
	\end{align*}
	Then the summation over $\mu$ in equation \eqref{eqn:main lem step 1} can be taken,
	and equation \eqref{eqn:mu,g-Qg+ ep,mu^t} can be reduced to
	\begin{align*}
		&\sum_{\mu'\in\mathcal{P}}
		\langle\mu'|
		Q^{{L_0}} \cdot
		\prod_{i=1}^L
		\big(\Gamma_+(y_{1,i}\cdot q^{\rho})^{\epsilon_{2,i}}
		\cdot
		\prod_{i=1}^L
		\Gamma_-(-x_{1,i}\cdot q^{\rho})^{-\epsilon_{1,i}}
		|\mu^{',t}\rangle\\
		=&\prod_{i,k=1}^L
		\frac{1}
		{M(-x_{1,i}y_{1,k};q)^{-\epsilon_{1,i}\epsilon_{2,k}}}\\
		&\quad\quad\quad\quad\quad\quad\cdot \sum_{\mu'\in\mathcal{P}}
		\langle\mu'|
		\prod_{i=1}^L
		\Gamma_-(-Qx_{1,i}\cdot q^{\rho})^{-\epsilon_{1,i}}
		\cdot Q^{{L_0}} \cdot
		\prod_{i=1}^L
		\Gamma_+(y_{1,i}\cdot q^{\rho})^{\epsilon_{2,i}}
		\cdot
		|\mu^{',t}\rangle.
	\end{align*}
	Next,
	we can insert the identity operator 
	$\sum_{\mu\in\mathcal{P}}|\mu^t\rangle\langle\mu^t|$
	behind $Q^{L_0}$,
	and then similarly obtain that,
	equation \eqref{eqn:mu,g-Qg+ ep,mu^t} reduces to
	\begin{align*}
		&\prod_{i,k=1}^L
		\prod_{j=0}^1
		\frac{1}
		{M\big((-1)^{j-1}Q^jx_{1,i}y_{1,k};q\big)^{(-1)^{j-1}\epsilon_{1,i}\epsilon_{2,k}}}\\
		&\quad\quad\quad\quad\quad\quad\cdot \sum_{\mu\in\mathcal{P}}
		\langle\mu|
		\prod_{i=1}^L
		\Gamma_-(-Qx_{1,i}\cdot q^{\rho})^{-\epsilon_{1,i}}
		\cdot Q^{{L_0}} \cdot
		\prod_{i=1}^L
		\Gamma_+(-Qy_{1,i}\cdot q^{\rho})^{-\epsilon_{2,i}}
		|\mu^t\rangle.
	\end{align*}
	Thus,
	by doing the above process again and again and using the fact that $|Q|<1$,
	the equation \eqref{eqn:mu,g-Qg+ ep,mu^t} is equal to
	\begin{align*}
		&\prod_{i,k=1}^L
		\prod_{j=0}^{\infty}
		\frac{1}
		{M\big((-1)^{j-1}Q^jx_{1,i}y_{1,k};q\big)^{(-1)^{j-1}\epsilon_{1,i}\epsilon_{2,k}}}
		\cdot \sum_{\mu\in\mathcal{P}}
		\langle\mu|Q^{{L_0}}|\mu^t\rangle.
	\end{align*}
	This lemma then follows from the following identity
	\begin{align*}
		\sum_{\mu\in\mathcal{P}}
		\langle\mu|
		Q^{{L_0}}
		|\mu^t\rangle
		=\sum_{\mu=\mu^t\in\mathcal{P}} Q^{|\mu|}
		=\prod_{n\geq1}(1+Q^{2n-1}).
	\end{align*}
	$\Box$

	\begin{cor}\label{cor:mu,g--Qg++ ep,mu^t}
		We assume $|Q|<1$, then
		\begin{align}\label{eqn:mu,g--Qg++ ep,mu^t}
			\sum_{\mu\in\mathcal{P}}
			\langle \mu|
			\prod_{i=1}^N
			\big(\Gamma_-(x_{1,i}\cdot q^{\rho})^{\epsilon_{1,i}}
			&\Gamma_-(x_{2,i}\cdot q^{\rho})^{\epsilon_{2,i}}\big)
			\cdot Q^{{L_0}} \cdot
			\prod_{i=1}^N 
			\big(\Gamma_+(y_{1,i}\cdot q^{\rho})^{\epsilon_{3,i}}
			\Gamma_+(y_{2,i}\cdot q^{\rho})^{\epsilon_{4,i}}\big)
			|\mu^t\rangle \nonumber\\
			&\quad\quad\quad
			=\prod_{n\geq1}(1+Q^{2n-1})
			\cdot \prod_{i,k=1}^N
			\prod_{j=0}^{\infty}
			h_{i,k,j},
		\end{align}
		where
		\begin{align*}
			h_{i,k,j}=
			\frac{M\big((-1)^{j-1}Q^jx_{1,i}y_{2,k};q\big)^{(-1)^{j}\epsilon_{1,i}\epsilon_{4,k}}
				\cdot M\big((-1)^{j-1}Q^jx_{2,i}y_{1,k};q\big)^{(-1)^{j}\epsilon_{2,i}\epsilon_{3,k}}}
			{M\big((-1)^{j-1}Q^jx_{1,i}y_{1,k};q\big)^{(-1)^{j-1}\epsilon_{1,i}\epsilon_{3,k}}
				\cdot M\big((-1)^{j-1}Q^jx_{2,i}y_{2,k};q\big)^{(-1)^{j-1}\epsilon_{2,i}\epsilon_{4,k}}}.
		\end{align*}
	\end{cor}
	{\bf Proof:}
	Apply Lemma \ref{lem:mu,g-Qg+ ep,mu^t} to the case $L=2N$.
	$\Box$
	\\\ 
	
	We will show that both of the two sides of equation \eqref{eqn:main} are equal to the partition function $Z_N$ (defined in equation \eqref{eqn:def Z_N}) up to a multiplicative factor $\prod_{i=1}^N M(Q_{1,i};q)$ in the following two Propositions \ref{pro:Z_N as inf prod} and \ref{pro:Z_N as inf sum}.
	Then Theorem \ref{thm:main} follows directly from them.
	
	\begin{prop}\label{pro:Z_N as inf prod}
		For any positive integer $N$,
		we have the following product formula for $Z_N$ as an element of the ring $\mathbb{C}[\![Q_{1,i},Q_{2,i},q^{-1}]\!]$.
		\begin{align}
			Z_N
			=\begin{cases}
				\frac{\prod_{i=1}^N M(Q_{1,i};q)}
				{\prod_{n\geq1}
					(1-\prod_{i=1}^NQ_{1,i}^nQ_{2,i}^n)}
				\cdot \prod_{1\leq i<k\leq N} f_{k,i,-1}
				\cdot\prod_{i,k=1}^N
				\prod_{j=0}^{\infty} f_{i,k,j}, \text{\ if\ }N\text{\ is\ even},\\
				\frac{\prod_{i=1}^N M(Q_{1,i};q)}
				{\prod_{n\geq1}
					(1+\prod_{i=1}^NQ_{1,i}^{2n-1}Q_{2,i}^{2n-1})^{-1}}
				\cdot \prod_{1\leq i<k\leq N} g_{k,i,-1}
				\cdot\prod_{i,k=1}^N
				\prod_{j=0}^{\infty} g_{i,k,j}, \text{\ if\ }N\text{\ is\ odd}.\\
			\end{cases}
		\end{align}
		We repeat the notations defined in Theorem \ref{thm:main} here for convenience.
		Denote by $M(z;q)=\prod_{n=1}^\infty (1-zq^{-n})^n$,
		\begin{align*}
			f_{i,k,j}=&\frac{M(Q_{1,i} Q_{1,k} a_{i,k,j};q)^{\epsilon_i \epsilon_k}
				\cdot M(a_{i,k,j};q)^{\epsilon_i \epsilon_k}}
			{M(Q_{1,k} a_{i,k,j};q)^{\epsilon_i \epsilon_k}
				\cdot M(Q_{1,i} a_{i,k,j};q)^{\epsilon_i \epsilon_k}},\\
			g_{i,k,j}=&\frac{M\big((-1)^{j-1}Q_{1,i} Q_{1,k} a_{i,k,j};q\big)^{(-1)^{j-1}\epsilon_i \epsilon_k}
				\cdot M\big((-1)^{j-1}a_{i,k,j};q\big)^{(-1)^{j-1}\epsilon_i \epsilon_k}}
			{M\big((-1)^{j-1}Q_{1,k} a_{i,k,j};q\big)^{(-1)^{j-1}\epsilon_i \epsilon_k}
				\cdot M\big((-1)^{j-1}Q_{1,i} a_{i,k,j};q\big)^{(-1)^{j-1}\epsilon_i \epsilon_k}},\\
			a_{i,k,j}=&\epsilon_i \epsilon_k\cdot
			Q_{2,k} \cdot \prod_{n=1}^{i-1}Q_{1,n}Q_{2,n}
			\cdot \prod_{n=k+1}^NQ_{1,n}Q_{2,n}
			\cdot \prod_{n=1}^N(Q_{1,n}Q_{2,n})^{j},
		\end{align*}
		and $\epsilon_i=\delta_{i,odd}=1$ if $i$ is odd, otherwise it is $-1$.
	\end{prop}
	{\bf Proof:}
	First,
	by the rotation symmetry \eqref{eqn:rotation symm C} of the topological vertex,
	the partition function $Z_N$ defined in equation \eqref{eqn:def Z_N} can be written as,
	\begin{align*}
		Z_N(Q_{1,i},Q_{2,i};q)
		=\sum_{\mu^1,\nu^1,...,\mu^N,\nu^N \in\mathcal{P}}
		\prod_{i=1}^N(-Q_{1,i})^{|\mu^i|}(-Q_{2,i})^{|\nu^i|}
		C_{\mu^{i,t},\nu^i,\emptyset}(q) C_{\mu^{i+1},\nu^{i},\emptyset}(q).
	\end{align*}
	Then, by the formula \eqref{eqn:clm0=<>} for the special case of the topological vertex via the vacuum expectation value,
	we have
	\begin{align}\label{eqn:Z_N first vev}
		Z_N=&\sum_{\mu^1,\nu^1,...,\mu^N,\nu^N}
		\prod_{i=1}^N(-Q_{1,i})^{|\mu^i|}(-Q_{2,i})^{|\nu^i|}
		\langle\mu^i|
		\Gamma_-(q^{\rho})\Gamma_+(q^{\rho})
		|\nu^{i}\rangle
		\cdot
		\langle\mu^{i+1,t}|
		\Gamma_-(q^{\rho})\Gamma_+(q^{\rho})
		|\nu^{i}\rangle.
	\end{align}
	To deal with the right-hand side of the above equation \eqref{eqn:Z_N first vev},
	we take the $w-$transform of the penultimate factor, i.e., we use the following identity (see equation \eqref{eqn:mu,mu^t -1})
	\begin{align*}
		\langle\mu^i|
		\Gamma_-(q^{\rho})\Gamma_+(q^{\rho})
		|\nu^{i}\rangle
		=\langle\mu^{i,t}|
		\Gamma_-(-q^{\rho})^{-1}\Gamma_+(-q^{\rho})^{-1}
		|\nu^{i,t}\rangle.
	\end{align*}
	For the last factor in above equation \eqref{eqn:Z_N first vev},
	we take a dual of it, i.e., we use
	\begin{align*}
		\langle\mu^{i+1,t}|
		\Gamma_-(q^{\rho})\Gamma_+(q^{\rho})
		|\nu^{i}\rangle
		=\langle\nu^{i}|
		\Gamma_-(q^{\rho})\Gamma_+(q^{\rho})
		|\mu^{i+1,t}\rangle,
	\end{align*}
	and then shift the index $i\mapsto i-1$.
	The result is that,
	after summing over all $\mu^i, 1\leq i \leq N$,
	equation \eqref{eqn:Z_N first vev} reduces to
	\begin{align*}
		Z_N=&\sum_{\nu^1,...,\nu^N}
		\prod_{i=1}^N
		\langle\nu^{i-1}|
		\Gamma_-(q^{\rho})\Gamma_+(q^{\rho})
		\cdot (-Q_{1,i})^{L_0} \cdot
		\Gamma_-(-q^{\rho})^{-1}
		\Gamma_+(-q^{\rho})^{-1}
		(-Q_{2,i})^{L_0}|\nu^{i,t}\rangle,
	\end{align*}
	where we use the notation $\nu^{0}=\nu^N$.
	Thus,
	by the commutation relations \eqref{eqn:comm g+-} and \eqref{eqn:comm gL0},
	we can adjust the orders of $\Gamma_\pm(\cdot)$ and $Q^{L_0}$ to obtain 
	\begin{align*}
		Z_N=\sum_{\nu^1,...,\nu^N}
		\prod_{i=1}^N
		M(Q_{1,i};q)\cdot&
		\langle\nu^{i-1}|
		\Gamma_-(q^{\rho})
		\Gamma_-(Q_{1,i}q^{\rho})^{-1}\\
		&\quad\quad\cdot (Q_{1,i}Q_{2,i})^{L_0}
		\cdot \Gamma_+(Q_{1,i}Q_{2,i}q^{\rho})
		\Gamma_+(Q_{2,i}q^{\rho})^{-1}
		|\nu^{i,t}\rangle.
	\end{align*}
	We can see that,
	since the vacuum expectation value in the above equation is of the form
	\[\langle\nu^{i-1}|\cdots|\nu^{i,t}\rangle.\]
	we cannot directly sum over $\nu^i$.
	There are two cases.
	
	{\bf Case 1: } $N$ is an even positive integer.
	Then by taking the $w$-transform to the terms labeled by even integer $i$,
	we obtain,
	\begin{align*}
		&Z_N=
		\prod_{i=1}^N M(Q_{1,i};q)\\
		&\cdot\sum_{\nu^1,...,\nu^N}
		\prod_{i=1,odd}^N
		\langle\nu^{i-1}|
		\Gamma_-(q^{\rho})
		\Gamma_-(Q_{1,i}q^{\rho})^{-1}
		\cdot
		(Q_{1,i}Q_{2,i})^{L_0}\cdot
		\Gamma_+(Q_{1,i}Q_{2,i}q^{\rho})
		\Gamma_+(Q_{2,i}q^{\rho})^{-1}
		|\nu^{i,t}\rangle\\
		&\cdot
		\prod_{i=1,even}^N
		\langle\nu^{i-1,t}|
		\Gamma_-(-q^{\rho})^{-1}
		\Gamma_-(-Q_{1,i}q^{\rho})
		\cdot
		(Q_{1,i}Q_{2,i})^{L_0}\cdot
		\Gamma_+(-Q_{1,i}Q_{2,i}q^{\rho})^{-1}
		\Gamma_+(-Q_{2,i}q^{\rho})
		|\nu^{i}\rangle.
	\end{align*}
	Thus,
	the sum over $\nu^i$ for $i=1,2,...,N-1$ can be performed,
	and then we have
	\begin{align*}
		Z_N
		=\prod_{i=1}^N M(Q_{1,i};q)
		\cdot&\sum_{\nu^N\in\mathcal{P}}
		\langle\nu^{N}|
		\prod_{i=1}^N\Big(
		\Gamma_-(\epsilon_iq^{\rho})^{\epsilon_i}
		\Gamma_-(\epsilon_iQ_{1,i}q^{\rho})^{-\epsilon_i}\\
		&\quad\quad\quad\cdot
		(Q_{1,i}Q_{2,i})^{L_0}\cdot
		\Gamma_+(\epsilon_iQ_{1,i}Q_{2,i}q^{\rho})^{\epsilon_i}
		\Gamma_+(\epsilon_iQ_{2,i}q^{\rho})^{-\epsilon_i}\Big)
		|\nu^{N}\rangle,
	\end{align*}
	where $\epsilon_i=\delta_{i,odd}-\delta_{i,even}=1$ if $i$ is odd, otherwise it is $-1$.
	
	Using the commutation relations \eqref{eqn:comm g+-} and \eqref{eqn:comm gL0} again,
	we change the orders of $\Gamma_\pm(\cdot)$ and $Q^{L_0}$ to obtain
	\begin{align}\label{eqn:Z_N final even N}
		Z_N&
		=\prod_{i=1}^N M(Q_{1,i};q)
		\cdot \prod_{1\leq i<k\leq N}
		f_{k,i,-1} \nonumber\\
		&\cdot\sum_{\nu^N\in\mathcal{P}}
		\langle\nu^N|
		\prod_{i=1}^N
		\Big(\Gamma_-(\epsilon_i
		\prod_{j=1}^{i-1}Q_{1,j}Q_{2,j}
		\cdot q^{\rho})^{\epsilon_i}
		\Gamma_-(\epsilon_i Q_{1,i}
		\prod_{j=1}^{i-1}Q_{1,j}Q_{2,j}
		\cdot q^{\rho})^{-\epsilon_i}\Big) \nonumber\\
		&\quad\cdot
		(\prod_{i=1}^N Q_{1,i}Q_{2,i})^{L_0}\cdot
		\prod_{i=1}^N
		\Big( \Gamma_+(\epsilon_i
		\prod_{j=i}^{N}Q_{1,j}Q_{2,j}
		\cdot q^{\rho})^{\epsilon_i}
		\Gamma_+(\epsilon_i Q_{2,i}
		\prod_{j=i+1}^{N}Q_{1,j}Q_{2,j}
		\cdot q^{\rho})^{-\epsilon_i}\Big)
		|\nu^N\rangle,
	\end{align}
	where $f_{i,k,j}$ is defined in equation \eqref{eqn:def f}
	and for convenience,
	we rewrite it again here for the case used in the above equation \eqref{eqn:Z_N final even N}.
	That is to say,
	for $i<k$,
	denote by
	\begin{align*}
		f_{k,i,-1}
		=\frac{M(\epsilon_i\epsilon_kQ_{1,k}
			\prod_{j=i}^{k-1}Q_{1,j}Q_{2,j} ;q)^{\epsilon_i\epsilon_k}
			\cdot M(\epsilon_i\epsilon_kQ_{2,i}
			\prod_{j=i+1}^{k-1}Q_{1,j}Q_{2,j} ;q)^{\epsilon_i\epsilon_k}}
		{M(\epsilon_i\epsilon_k
			\prod_{j=i}^{k-1}Q_{1,j}Q_{2,j} ;q)^{\epsilon_i\epsilon_k}
			\cdot M(\epsilon_i\epsilon_k
			Q_{2,i}Q_{1,k}
			\prod_{j=i+1}^{k-1}Q_{1,j}Q_{2,j} ;q)^{\epsilon_i\epsilon_k}}.
	\end{align*}
	As a consequence,
	the Corollary \ref{cor:mu,g--Qg++ ep,mu} for the case,
	\begin{align}\label{eqn:Q ep for Z_N}
		Q=\prod_{i=1}^N Q_{1,i}Q_{2,i},
		\quad\quad\epsilon_{k,i}=(-1)^{k-1}\epsilon_{i}=(-1)^{k-1}(\delta_{i,odd}-\delta_{i,even}),
		\quad 1\leq k \leq4,
		\quad 1\leq i \leq N,
	\end{align}
	and
	\begin{align}\label{eqn:xy for Z_N}
		\begin{split}
			&x_{1,i}=\epsilon_{i}
			\prod_{j=1}^{i-1}Q_{1,j}Q_{2,j},
			\quad x_{2,i}=\epsilon_{i}
			Q_{1,i}\prod_{j=1}^{i-1}Q_{1,j}Q_{2,j},\\
			&y_{1,i}=\epsilon_i \prod_{j=i}^NQ_{1,j}Q_{2,j},
			\quad y_{2,i}=\epsilon_{i}Q_{2,i} \prod_{j=i+1}^NQ_{1,j}Q_{2,j},
			\quad 1\leq i\leq N,
		\end{split}
	\end{align}
	can be applied to compute equation \eqref{eqn:Z_N final even N}.
	Then, the even $N$ case of this proposition is proved.
	
	{\bf Case 2: } $N$ is an odd positive integer.
	Similar to Case 1,
	we first take the $w$-transform to the terms labeled by even integer $i$,
	and then obtain
	\begin{align*}
		&Z_N=
		\prod_{i=1}^N M(Q_{1,i};q)\\
		&\cdot\sum_{\nu^1,...,\nu^N}
		\prod_{i=1,odd}^N
		\langle\nu^{i-1}|
		\Gamma_-(q^{\rho})
		\Gamma_-(Q_{1,i}q^{\rho})^{-1}
		\cdot
		(Q_{1,i}Q_{2,i})^{L_0}\cdot
		\Gamma_+(Q_{1,i}Q_{2,i}q^{\rho})
		\Gamma_+(Q_{2,i}q^{\rho})^{-1}
		|\nu^{i,t}\rangle\\
		&\cdot
		\prod_{i=1,even}^N
		\langle\nu^{i-1,t}|
		\Gamma_-(-q^{\rho})^{-1}
		\Gamma_-(-Q_{1,i}q^{\rho})
		\cdot
		(Q_{1,i}Q_{2,i})^{L_0}\cdot
		\Gamma_+(-Q_{1,i}Q_{2,i}q^{\rho})^{-1}
		\Gamma_+(-Q_{2,i}q^{\rho})
		|\nu^{i}\rangle.
	\end{align*}
	After taking the summation over $\nu^i$ for $i=1,2,...,N-1$ in the above equation,
	we have
	\begin{align*}
		Z_N
		=\prod_{i=1}^N M(Q_{1,i};q)
		&\cdot\sum_{\nu^N\in\mathcal{P}}
		\langle\nu^{N}|
		\prod_{i=1}^N\Big(
		\Gamma_-(\epsilon_iq^{\rho})^{\epsilon_i}
		\Gamma_-(\epsilon_iQ_{1,i}q^{\rho})^{-\epsilon_i}\\
		&\quad\quad\quad\cdot
		(Q_{1,i}Q_{2,i})^{L_0}\cdot
		\Gamma_+(\epsilon_iQ_{1,i}Q_{2,i}q^{\rho})^{\epsilon_i}
		\Gamma_+(\epsilon_iQ_{2,i}q^{\rho})^{-\epsilon_i}\Big)
		|\nu^{N,t}\rangle.
	\end{align*}
	Note that,
	the final expression for $Z_N$ is of the form
	\[\sum_{\mu} \langle\mu| \cdots |\mu^t\rangle,\]
	which is different from Case 1.
	However,
	the methods for dealing with such formulas are similar.
	Using the commutation relations \eqref{eqn:comm g+-} and \eqref{eqn:comm gL0},
	we change the positions of $\Gamma_\pm(\cdot)$ and $Q^{L_0}$.
	Then
	\begin{small}
		\begin{align}\label{eqn:Z_N final odd N}
			\begin{split}
				Z_N
				=&\prod_{i=1}^N M(Q_{1,i};q)
				\cdot \prod_{1\leq i<k\leq N}
				g_{k,i,-1} \\
				&\cdot\sum_{\nu^N\in\mathcal{P}}
				\langle\nu^N|
				\prod_{i=1}^N
				\Big(\Gamma_-(\epsilon_i
				\prod_{j=1}^{i-1}Q_{1,j}Q_{2,j}
				\cdot q^{\rho})^{\epsilon_i}
				\Gamma_-(\epsilon_i Q_{1,i}
				\prod_{j=1}^{i-1}Q_{1,j}Q_{2,j}
				\cdot q^{\rho})^{-\epsilon_i}\Big)\\
				&\quad\cdot
				(\prod_{i=1}^N Q_{1,i}Q_{2,i})^{L_0}\cdot
				\prod_{i=1}^N
				\Big( \Gamma_+(\epsilon_i
				\prod_{j=i}^{N}Q_{1,j}Q_{2,j}
				\cdot q^{\rho})^{\epsilon_i}
				\Gamma_+(\epsilon_i Q_{2,i}
				\prod_{j=i+1}^{N}Q_{1,j}Q_{2,j}
				\cdot q^{\rho})^{-\epsilon_i}\Big)
				|\nu^{N,t}\rangle,
			\end{split}
		\end{align}
	\end{small}
	where $g_{i,k,j}$ is defined in equation \eqref{eqn:def g}.
	Notice that,
	when $j$ is an odd integer, $g_{i,k,j}=f_{i,k,j}$.
	Thus,
	Corollary \ref{cor:mu,g--Qg++ ep,mu^t} can be applied to calculate the above equation \eqref{eqn:Z_N final odd N}.
	The parameters in Corollary \ref{cor:mu,g--Qg++ ep,mu^t} are chosen according to equations \eqref{eqn:Q ep for Z_N} and \eqref{eqn:xy for Z_N}.
	This proposition is thus proved.
	$\Box$

	\subsection{Infinite sum formula for $Z_N$}
	
	To obtain the left-hand side of equation \eqref{eqn:main},
	we first need the following combinatorial lemma.
	A similar result without proof
	(whatever, some indexes in their formula should also be adjusted)
	can be seen in Section 3 in \cite{INRS}.
	A $(q,t)-$version can be seen in Proposition 2.14 in \cite{AFHKSY} and \cite{CNO}.
	\begin{lem}\label{lem:inf/inf=finite}
		For any partitions $\mu$ and $\nu$,
		we have
		\begin{align}\label{eqn:inf/inf=finite}
			\sum_{\lambda}
			z^{|\lambda|}
			s_{\lambda}(q^{\mu+\rho})
			s_{\lambda^t}(q^{\nu^t+\rho})
			=&\prod_{(j,k)\in\mu}(1+zq^{\mu_j+\nu^t_k-j-k+1})
			\cdot\prod_{(j,k)\in\nu}(1+zq^{-\mu^t_k-\nu_j+j+k-1})\\
			&\quad\quad\quad\quad\quad
			\cdot\prod_{j,k=1}^\infty (1+zq^{-j-k+1}).
		\end{align}
	\end{lem}
	{\bf Proof:}
	Denote by $W(\mu,\nu)$ the left-hand side of equation \eqref{eqn:inf/inf=finite}.
	Recall that by equations \eqref{eqn:gamma- action}, \eqref{eqn:gamma+ action} and \eqref{eqn:mu,mu^t -1}, we have
	\begin{align*}
		\langle0|\Gamma_+(\bm x)
		=\sum_\lambda s_{\lambda}(\bm x) \langle\lambda|
		\quad\text{and}\quad
		\langle\lambda|\Gamma_-(-\bm x)^{-1}|0\rangle
		=\langle\lambda^t|\Gamma_-(\bm x)|0\rangle=s_{\lambda^t}(\bm x).
	\end{align*}
	Then $W(\mu,\nu)$ can be represented by the following vacuum expectation value
	\begin{align}\label{eqn:W vev formula}
		W(\mu,\nu)
		=\langle0|
		\Gamma_+(q^{\mu+\rho})
		\cdot z^{L_0} \cdot
		\Gamma_-(-q^{\nu^t+\rho})^{-1}
		|0\rangle.
	\end{align}
	We will prove this lemma by induction on the length of $\mu$.
	The author used a similar method to study the partition function of certain plane partitions (see the proof in Theorem 1.3 in \cite{Y23}).
	
	First,
	if $l(\mu)=0$,
	i.e. $\mu=\emptyset$,
	\begin{align*}
		W(\emptyset,\nu)
		=\langle0|
		\Gamma_+(q^{\rho})
		\cdot z^{L_0} \cdot
		\Gamma_-(-q^{\nu^t+\rho})^{-1}
		|0\rangle
		=\prod_{j,k=1}^\infty
		(1+zq^{\nu^{t}_k-j-k+1})
	\end{align*}
	by commutation relations \eqref{eqn:comm g+-} and \eqref{eqn:comm gL0}.
	For each fixed $k$,
	we divide the product over $\{j=1,2,...\}$ to two subsets
	$\{j=1,...,\nu^t_k\}$ and $\{j=\nu^t_k+1,\nu^t_k+2,...\}$.
	The result is
	\begin{align*}
		W(\emptyset,\nu)
		=\prod_{k=1}^{l(\nu^t)} \prod_{j=1}^{\nu^t_k}
		(1+zq^{\nu^t_k-j-k+1})
		\cdot\prod_{k=1}^\infty \prod_{j=\nu^t_{k}+1}^{\infty} (1+zq^{\nu^t_k-j-k+1}).
	\end{align*}
	By letting $j\mapsto\nu^t_k+1-j, k\mapsto \nu_j+1-k$,
	the first factor of the right-hand side of the above equation reduces to
	\[\prod_{(j,k)\in\nu} (1+zq^{-\nu_j+j+k-1}).\]
	By letting $j\mapsto j+\nu^t_k$,
	the second factor of the above equation reduces to
	$\prod_{j,k=1}^\infty (1+zq^{-j-k+1})$.
	Thus, the $\mu=\emptyset$ case of this lemma is proved.
	
	From now on,
	we assume $l(\mu)=l>0$ and this lemma already holds for any $\tilde{\mu}$ satisfying $l(\tilde{\mu})<l(\mu)$.
	Especially,
	this lemma holds for $\mu'=(\mu_1,...,\mu_{l(\mu)-1})$ and any $\nu$.
	Thus,
	by equation \eqref{eqn:W vev formula},
	we have
	\begin{align*}
		W(\mu,\nu)
		=&\langle0|
		\prod_{j=1\atop j\neq l}^\infty
		\Gamma_+(q^{\mu_j-j+\half})
		\cdot \Gamma_+(q^{\mu_l-l+\half})
		\cdot z^{L_0} \cdot
		\Gamma_-(-q^{\nu^t+\rho})^{-1}
		|0\rangle.
	\end{align*}
	We replace the term $\Gamma_+(q^{\mu_l-l+\half})$ appeared in the right-hand side of above equation by
	\begin{align*}
		\Gamma_+(q^{-l+\half})
		\cdot \Gamma_+(q^{-l+\half})^{-1}\Gamma_+(q^{\mu_l-l+\half})
	\end{align*}
	and then move $\Gamma_+(q^{-l+\half})^{-1}\Gamma_+(q^{\mu_l-l+\half})$
	to the rightmost side.
	Since $\Gamma_+(\cdot)$ preserves the vacuum vector $|0\rangle$,
	we obtain
	\begin{align*}
		W(\mu,\nu)
		=&W(\mu',\nu)
		\cdot \prod_{k=1}^\infty \frac{(1+zq^{\mu_l+\nu^t_k-l-k+1})}
		{(1+zq^{\nu^t_k-l-k+1})}.
	\end{align*}
	Thus, by induction process,
	to prove this lemma,
	it remains to show
	\begin{align}\label{eqn:only need1}
		\prod_{k=1}^{\mu_l}(1+zq^{\mu_l+\nu^t_k-l-k+1})
		\cdot\prod_{(j,k)\in\nu}
		\frac{(1+zq^{-\mu^t_k-\nu_j+j+k-1})}
		{(1+zq^{-\mu^{',t}_k-\nu_j+j+k-1})}
		=\prod_{k=1}^\infty \frac{(1+zq^{\mu_l+\nu^t_k-l-k+1})}
		{(1+zq^{\nu^t_k-l-k+1})}.
	\end{align}
	Notice that there are three combinatorial facts,
	which can simply above equation \eqref{eqn:only need1}.\\
	i) the first factor in the left-hand side of the above equation is a part of the numerator of the right-hand side of the above equation.\\
	ii) $\mu^t_k=\mu^{',t}_k+1=l$ when $1\leq k\leq\mu_l$,
	and $\mu^t_k=\mu^{',t}_k$ when $k>\mu_l$.
	This result can simplify the second factor in the left-hand side of the above equation.\\
	iii) $\nu^t_k=0$ when $k>\nu_1$.
	This result can be used to simplify the right-hand side of the above equation.
	
	As a consequence,
	above equation \eqref{eqn:only need1} is equivalent to the following combinatorial identity:
	\begin{align}\label{eqn:only need2}
		\prod_{j=1\atop\nu_j\geq\mu_l}^{l(\nu)}
		\frac{(1+zq^{-l-\nu_j+j})}
		{(1+zq^{-l-\nu_j+j+\mu_l})}
		\cdot \prod_{j=1\atop\nu_j<\mu_l}^{l(\nu)}
		\frac{(1+zq^{-l-\nu_j+j})}
		{(1+zq^{-l+j})}
		=\frac{\prod_{k=\mu_l+1}^{\nu_1+\mu_l} (1+zq^{\mu_l+\nu^t_k-l-k+1})}
		{\prod_{k=1}^{\nu_1} (1+zq^{\nu^t_k-l-k+1})}.
	\end{align}
	We will then prove above equation \eqref{eqn:only need2} by induction on the length of $\nu$,
	and thus, this lemma is proved.
	
	The initial case is $l(\nu)=0$, i.e. $\nu=\emptyset$.
	Then, both sides of equation \eqref{eqn:only need2} are obviously equal to 1.
	For fixed $l, \mu_l$ and $\nu$,
	we assume that $l(\nu)=n>0$ and equation \eqref{eqn:only need2} holds for any $\tilde{\nu}$ such that $l(\tilde{\nu})<n$.
	In particular,
	it holds for $\nu'=(\nu_1,...,\nu_{n-1})$.
	Denote by $D(\nu)$ and $E(\nu)$ the left and the right-hand side of equation \eqref{eqn:only need2} respectively.
	Thus, we can assume $D(\nu')=E(\nu')$ by induction.
	Then, by definition,
	\begin{align}\label{eqn:D recursion}
		D(\nu)
		=D(\nu')
		\cdot\begin{cases}
			\frac{(1+zq^{-l-\nu_n+n})}
			{(1+zq^{-l-\nu_n+n+\mu_l})}, & \text{\ if\ }\nu_n\geq\mu_l,\\
			\frac{(1+zq^{-l-\nu_n+n})}
			{(1+zq^{-l+n})}, & \text{\ if\ }\nu_n<\mu_l.
		\end{cases}
	\end{align}
	On the other hand,
	since $\nu^t_k=\nu^{',t}_k+1=n$ when $1\leq k\leq\nu_n$,
	and $\nu^t_k=\nu^{',t}_k$ when $k>\nu_n$,
	$E(\nu)$ has the following equation,
	\begin{align}\label{eqn:E recursion}
		E(\nu)
		=E(\nu')
		\cdot \frac{(1+zq^{n-l-\nu_n})}{(1+zq^{n-l})}
		\cdot\begin{cases}
			1, &\text{\ if\ }\nu_n\leq\mu_l,\\
			\frac{(1+zq^{n-l})}{(1+zq^{\mu_l+n-l-\nu_n})} , & \text{\ if\ }\nu_n>\mu_l.
		\end{cases}
	\end{align}
	By comparing equations \eqref{eqn:D recursion} and \eqref{eqn:E recursion},
	we now know that $D(\nu)=E(\nu)$ for any $\nu$ by induction.
	
	Recall that the equation \eqref{eqn:only need1} follows from equation \eqref{eqn:only need2},
	and thus this lemma is proved.
	$\Box$

	\begin{prop}\label{pro:Z_N as inf sum}
		We have
		\begin{align}
			\begin{split}
				Z_N=&\prod_{i=1}^N M(Q_{1,i};q)
				\cdot \sum_{\nu^1,...,\nu^N \in\mathcal{P}}
				\prod_{i=1}^N(-Q_{2,i})^{|\nu^i|}
				q^{\kappa(\nu^i)-\|\nu^{i}\|^2}\\
				&\quad\quad\quad\cdot
				\prod_{(j,k)\in\nu^{i}}
				\frac{(1-Q_{1,i+1}q^{\nu^{i,t}_k+\nu^{i+1,t}_j-j-k+1})
					(1-Q_{1,i}q^{-\nu^{i-1}_k-\nu^{i}_j+j+k-1})}
				{(1-q^{-h(j,k)})^2}.
			\end{split}
		\end{align}
	\end{prop}
	{\bf Proof:}
	By the definition of $Z_N$ in equation \eqref{eqn:def Z_N}
	and the Schur expansion \eqref{eqn:def C} of the topological vertex,
	we have
	\begin{align*}
		Z_N=\sum_{\mu^1,\nu^1,...,\mu^N,\nu^N}
		\prod_{i=1}^N(-Q_{1,i})^{|\mu^i|}(-Q_{2,i})^{|\nu^i|}
		q^{\kappa(\nu^i)}
		s_{\nu^{i,t}}(q^\rho) s_{\mu^{i,t}}(q^{\nu^{i,t}+\rho})
		\cdot s_{\nu^{i,t}}(q^\rho) s_{\mu^{i+1}}(q^{\nu^{i,t}+\rho}).
	\end{align*}
	Thus,
	by shifting the index $i\mapsto i-1$ to the last two factors in the above equation,
	we obtain that,
	$Z_N$ is equal to
	\begin{align*}
		\sum_{\mu^1,\nu^1,...,\mu^N,\nu^N}
		\prod_{i=1}^N(-Q_{1,i})^{|\mu^i|}(-Q_{2,i})^{|\nu^i|}
		q^{\kappa(\nu^i)}
		s_{\nu^{i,t}}(q^\rho) s_{\mu^{i,t}}(q^{\nu^{i,t}+\rho})
		\cdot s_{\nu^{i-1,t}}(q^\rho) s_{\mu^{i}}(q^{\nu^{i-1,t}+\rho}).
	\end{align*}
	Then the Lemma \ref{lem:inf/inf=finite} can be applied to take the summation over $\mu^i$ in the above equation.
	The result is
	\begin{align*}
		Z_N
		=&\prod_{i=1}^N M(Q_{1,i};q)
		\cdot \sum_{\nu^1,...,\nu^N \in\mathcal{P}}
		\prod_{i=1}^N(-Q_{2,i})^{|\nu^i|}
		q^{\kappa(\nu^i)}
		s_{\nu^{i,t}}(q^\rho) s_{\nu^{i-1,t}}(q^\rho)\\
		&\quad\quad\quad
		\cdot\prod_{(j,k)\in\nu^{i-1,t}}
		(1-Q_{1,i}q^{\nu^{i-1,t}_j+\nu^{i,t}_k-j-k+1})
		\prod_{(j,k)\in\nu^{i}}
		(1-Q_{1,i}q^{-\nu^{i-1}_k-\nu^{i}_j+j+k-1}).
	\end{align*}
	Finally,
	the evaluation formula \eqref{eqn:schur q^rho} for the Schur functions can be applied to the above equation,
	and then
	\begin{align*}
		Z_N
		=&\prod_{i=1}^N M(Q_{1,i};q)
		\cdot \sum_{\nu^1,...,\nu^N \in\mathcal{P}}
		\prod_{i=1}^N(-Q_{2,i})^{|\nu^i|}
		q^{\kappa(\nu^i)-\|\nu^{i-1}\|^2/2-\|\nu^{i}\|^2/2}\\
		&\quad\quad\cdot
		\prod_{(j,k)\in\nu^{i-1,t}}
		\frac{(1-Q_{1,i}q^{\nu^{i-1,t}_j+\nu^{i,t}_k-j-k+1})}
		{(1-q^{-h(j,k)})}
		\cdot
		\prod_{(j,k)\in\nu^{i}}
		\frac{(1-Q_{1,i}q^{-\nu^{i-1}_k-\nu^{i}_j+j+k-1})}
		{(1-q^{-h(j,k)})}.
	\end{align*}
	By shifting the index $i\mapsto i+1$ for the penultimate factor in the above equation,
	this proposition is thus proved.
	$\Box$

	\subsection{Application of Theorem \ref{thm:main}}
	\label{sec:app of main thm}
	Inspired by \cite{INRS,NO06},
	we make the following change of variables
	\begin{align}\label{eqn:change beta}
		Q_{1,i}=e^{\beta t_i},
		\quad Q_{2,i}=s_i,
		\quad q=e^{\beta},
	\end{align}
	and then take $\beta\rightarrow0$ for equation \eqref{eqn:main}.
	The result is a much simpler equality that connects an infinite sum and an infinite product.

	For convenience,
	we restate Proposition \ref{cor:main} here.
	
	\begin{prop}[=Proposition \ref{cor:main}]
		We have
		\begin{small}
			\begin{align}\label{eqn:main app}
				\begin{split}
					&\sum_{\nu^1,...,\nu^N \in\mathcal{P}}
					\prod_{i=1}^N (-s_i)^{|\nu^i|}
					\prod_{(j,k)\in\nu^{i}}
					\begin{small}
						\frac{\big(t_{i+1}+(\nu^{i,t}_k+\nu^{i+1,t}_j-j-k+1)\big)
							\big(t_i-(\nu^{i-1}_k+\nu^{i}_j-j-k+1)\big)
						}{h(j,k)^2}
					\end{small}\\
					&\ =\begin{cases}
						\frac{1}
						{\prod\limits_{n\geq1}
							(1-s^n)}
						\cdot \prod\limits_{1\leq i<k\leq N} (1-b_{k,i,-1})^{\epsilon_i\epsilon_kt_it_k}
						\cdot\prod_{i,k=1}^N
						\prod_{j=0}^{\infty} (1-b_{i,k,j})^{\epsilon_i\epsilon_kt_it_k}, \text{\ if\ }N\text{\ is\ even},\\
						\prod\limits_{n\geq1}
						(1+s^{2n-1})
						\cdot \prod\limits_{1\leq i<k\leq N} (1-b_{k,i,-1})^{\epsilon_i\epsilon_kt_it_k}
						\cdot\prod\limits_{i,k=1}^N
						\prod\limits_{j=0}^{\infty} \big(1+(-1)^{j}b_{i,k,j}\big)^{\epsilon_j\epsilon_i\epsilon_kt_it_k}, \text{otherwise},
					\end{cases}
				\end{split}
			\end{align}
		\end{small}
		where $t_{N+1}=t_1, s=\prod_{i=1}^Ns_i$ and
		\begin{align*}
			b_{i,k,j}=\epsilon_i\epsilon_k
			\cdot\prod_{n=1}^{i-1}s_n
			\cdot\prod_{n=k}^N s_n
			\cdot\prod_{n=1}^N (s_n)^j.
		\end{align*}
	\end{prop}
	{\bf Proof:}
	After the change of variables \eqref{eqn:change beta} and taking $\beta\rightarrow0$,
	the left-hand side of equation \eqref{eqn:main} becomes
	\begin{small}
		\begin{align*}
			\sum_{\nu^1,...,\nu^N \in\mathcal{P}}
			\prod_{i=1}^N (-s_i)^{|\nu^i|}
			\prod_{(j,k)\in\nu^{i}}
			\frac{\big(t_{i+1}+(\nu^{i,t}_k+\nu^{i+1,t}_j-j-k+1)\big)
				\big(t_i-(\nu^{i-1}_k+\nu^{i}_j-j-k+1)\big)
			}{h(j,k)^2}.
		\end{align*}
	\end{small}
	
	For the right-hand side of equation \eqref{eqn:main},
	we cannot directly take the limit $\beta\rightarrow0$
	since there will be infinitely many same terms, resulting in an indeterminate form  $\frac{\infty}{\infty}$.
	Thus, we first need to simplify it and then take the limit.
	
	Next,
	we assume $t_i\in\mathbb{Z}_+, 1\leq i\leq N$,
	and we will prove equation \eqref{eqn:main app} under this assumption below.
	This is sufficient to finish the proof of this corollary for general $t_i$.
	The reason is that,
	after expanding both sides of equation \eqref{eqn:main app} with respect to variables $s_i, 1\leq i\leq N$ in the region $\{|s_i|<1, 1\leq i\leq N\}$,
	the coefficients of each terms are polynomials in $\{t_i\}_{i=1}^N$ of finite orders.
	Thus,
	if equation \eqref{eqn:main app} holds for all $t_i\in\mathbb{Z}_+, 1\leq i\leq N$, it shows that these polynomials are the same,
	and this corollary for general $t_i$ is proved automatically.

	The first case is when $N$ is an even positive integer.
	We deal with each of the terms appearing in $f_{i,k,j}$.
	After the change of variables \eqref{eqn:change beta},
	we have
	\begin{align*}
		M(Q_{1,i}Q_{1,k} a_{i,k,j};q)
		=\prod_{n=1}^\infty (1-e^{(t_i+t_k)\beta} a_{i,k,j} e^{-n \beta})^{n}.
	\end{align*}
	By shifting the index $n\mapsto n+t_i+t_k$ (here we need the assumption $t_i\in\mathbb{Z}_+$),
	\begin{align*}
		M(Q_{1,i}Q_{1,k} a_{i,k,j}&;q)
		=\prod_{n=1-t_i-t_k}^\infty (1-a_{i,k,j} e^{-n \beta})^{n+t_i+t_k}\\
		=&\prod_{n=1-t_i-t_k}^0 (1-a_{i,k,j} e^{-n \beta})^{n+t_i+t_k}
		\cdot\prod_{n=1}^\infty (1-a_{i,k,j} e^{-n \beta})^{t_i+t_k}
		\cdot M(a_{i,k,j};q).
	\end{align*}
	By the same method,
	other terms that appearing in $f_{i,k,j}$ can be rewritten as
	\begin{align*}
		M(Q_{1,k} a_{i,k,j};q)
		=&\prod_{n=1-t_k}^0 (1-a_{i,k,j} e^{-n \beta})^{n+t_k}
		\cdot\prod_{n=1}^\infty (1-a_{i,k,j} e^{-n \beta})^{t_k}
		\cdot M(a_{i,k,j};q),\\
		M(Q_{1,i} a_{i,k,j};q)
		=&\prod_{n=1-t_i}^0 (1-a_{i,k,j} e^{-n \beta})^{n+t_i}
		\cdot\prod_{n=1}^\infty (1-a_{i,k,j} e^{-n \beta})^{t_i}
		\cdot M(a_{i,k,j};q).
	\end{align*}
	Thus,
	under the assumption that $t_i\in\mathbb{Z}_+$,
	we have
	\begin{align*}
		f_{i,k,j}
		=\frac{\prod_{n=1-t_i-t_k}^0 (1-a_{i,k,j} e^{-n \beta})^{(n+t_i+t_k)\epsilon_i\epsilon_k}}
		{\prod_{n=1-t_k}^0 (1-a_{i,k,j} e^{-n \beta})^{(n+t_k)\epsilon_i\epsilon_k}
			\cdot\prod_{n=1-t_i}^0 (1-a_{i,k,j} e^{-n \beta})^{(n+t_i)\epsilon_i\epsilon_k}}.
	\end{align*}
	Denote by
	\begin{align*}
		b_{i,k,j}=\epsilon_i\epsilon_k
		\cdot\prod_{n=1}^{i-1}s_n
		\cdot\prod_{n=k}^N s_n
		\cdot\prod_{n=1}^N (s_n)^j,
	\end{align*}
	which is the limit of $a_{i,k,j}$ as $\beta\rightarrow0$.
	By taking the limit $\beta\rightarrow0$,
	we obtain
	\begin{align*}
		f_{i,k,j}\rightarrow&
		\frac{\prod_{n=1-t_i-t_k}^0 (1-b_{i,k,j})^{(n+t_i+t_k)\epsilon_i\epsilon_k}}
		{\prod_{n=1-t_k}^0 (1-b_{i,k,j})^{(n+t_k)\epsilon_i\epsilon_k}
			\cdot\prod_{n=1-t_i}^0 (1-b_{i,k,j})^{(n+t_i)\epsilon_i\epsilon_k}}\\
		=&(1-b_{i,k,j})^{\epsilon_i\epsilon_kt_it_k}
	\end{align*}
	As a result,
	the limit of the right-hand side of equation \eqref{eqn:main} is
	\begin{align*}
		\frac{1}{\prod_{n\geq1}
			\big(1-\prod_{i=1}^N s_i^n\big)}
		\cdot \prod_{1\leq i<k\leq N} (1-b_{k,i,-1})^{\epsilon_i\epsilon_kt_it_k}
		\cdot\prod_{i,k=1}^N
		\prod_{j=0}^{\infty} (1-b_{i,k,j})^{\epsilon_i\epsilon_kt_it_k}.
	\end{align*}
	
	The second case is that: $N$ is an odd integer.
	Similarly, one can prove that
	\begin{align*}
		g_{i,k,j}\rightarrow&
		\frac{\prod_{n=1-t_i-t_k}^0 \big(1+(-1)^jb_{i,k,j}\big)^{\epsilon_j(n+t_i+t_k)\epsilon_i\epsilon_k}}
		{\prod_{n=1-t_k}^0 \big(1+(-1)^jb_{i,k,j}\big)^{\epsilon_j(n+t_k)\epsilon_i\epsilon_k}
			\cdot\prod_{n=1-t_i}^0 \big(1+(-1)^jb_{i,k,j}\big)^{\epsilon_j(n+t_i)\epsilon_i\epsilon_k}}\\
		=&\big(1+(-1)^jb_{i,k,j}\big)^{\epsilon_j\epsilon_i\epsilon_kt_it_k}
	\end{align*}
	and
	\begin{align*}
		\prod_{n\geq1}
		\big(1+\prod_{i=1}^NQ_{1,i}^{2n-1}Q_{2,i}^{2n-1}\big)
		\rightarrow
		\prod_{n\geq1}
		\big(1+\prod_{i=1}^Ns_i^{2n-1}\big).
	\end{align*}
	This proposition is thus proved.
	$\Box$

	The $N=1$ case of equation \eqref{eqn:main app} yields 
	\begin{small}
		\begin{align*}
			\sum_{\nu\in\mathcal{P}}
			s^{|\nu|}
			\prod_{(j,k)\in\nu}
			\frac{\big(t+l(j,k)+l(k,j)+1\big)\big(t-a(j,k)-a(k,j)-1\big)}{h(j,k)^2}
			=\prod\limits_{n\geq1}
			\frac{\big(1-s^{n}\big)^{(-1)^n t^2+1}}{(1-s^{2n})},
		\end{align*}
	\end{small}
	where $a(j,k), l(j,k)$ are the arm-length and leg-length explained in subsection \ref{sec:schur}.

	\section{Proof of the second family of Nekrasov--Okounkov type formulas}\label{sec:prove main2}
	In this section,
	we prove Theorem \ref{thm:main2} and Proposition \ref{cor:main2}.
	
	\subsection{Proof of Theorem \ref{thm:main2}}
	The method used in this subsection is similar to that used in proving Theorem \ref{thm:main}.
	We consider the following partition function
	\begin{align}\label{eqn:def tZ_N}
		\tilde{Z}_N=\sum_{\mu^1,\nu^1,...,\mu^N,\nu^N \in\mathcal{P}}
		\prod_{i=1}^N(-Q_{1,i})^{|\mu^i|}(-Q_{2,i})^{|\nu^i|}
		C_{\emptyset,\mu^{i,t},\nu^i}(q) C_{\emptyset,\mu^{i+1},\nu^{i,t}}(q),
	\end{align}
	where $\mu^{N+1}=\mu^1$.
	When $N=1, 2$,
	the above partition function \eqref{eqn:def tZ_N} was considered in \cite{INRS} to obtain Nekrasov--Okounkov type formulas.
	Thus, our results in this section are direct generalizations of the formulas proposed in \cite{INRS}.

	First,
	we need the following lemma.
	\begin{lem}\label{lem:mu,g--Qg++,mu}
		We assume $|Q|<1$, then
		\begin{align}\label{eqn:mu,g--Qg++,mu}
			&\sum_{\mu\in\mathcal{P}}
			\langle \mu|
			\prod_{i=1}^N
			\big(\Gamma_-(x_{1,i}\cdot q^{\rho})^{-1}
			\Gamma_-(x_{2,i}\cdot q^{\rho})\big)
			\cdot Q^{{L_0}} \cdot
			\prod_{i=1}^N 
			\big(\Gamma_+(y_{1,i}\cdot q^{\rho})^{-1}
			\Gamma_+(y_{2,i}\cdot q^{\rho})\big)
			|\mu\rangle\\
			&\quad\quad\quad\quad\quad\quad
			=\prod_{n\geq1}\frac{1}{(1-Q^n)}
			\cdot \prod_{i,k=1}^N
			\prod_{j=0}^{\infty}
			\frac{M(Q^j x_{1,i}y_{2,k};q)
				\cdot M(Q^j x_{2,i}y_{1,k};q)}
			{M(Q^j x_{1,i}y_{1,k};q)
				\cdot M(Q^j x_{2,i}y_{2,k};q)}.
		\end{align}
	\end{lem}
	{\bf Proof:}
	Just by applying the $L=2N$ and
	\begin{align*}
		\epsilon_{1,i}=\epsilon_{2,i}
		=\begin{cases}
			-1,&\text{\ if\ }1\leq i \leq N,\\
			1,&\text{\ if\ }N+1\leq i \leq 2N
		\end{cases}
	\end{align*}
	case of Lemma \ref{lem:mu,g-Qg+ ep,mu}.
	$\Box$
	\ \\

	In the following,
	we will show that both sides of equation \eqref{eqn:main2} are equal to the partition function $\tilde{Z}_N$ (defined in equation \eqref{eqn:def tZ_N}) up to the factor $\prod_{i=1}^N M(Q_{1,i};q)$.
	Then, Theorem \ref{thm:main2} follows directly.
	\begin{prop}
		For any positive integer $N$,
		we have the following product formula for $\tilde{Z}_N$ as an element of the ring $\mathbb{C}[\![Q_{1,i},Q_{2,2},q^{-1}]\!]$,
		\begin{align}
			\tilde{Z}_N
			=&\frac{\prod_{i=1}^N M(Q_{1,i};q)}
			{\prod_{n\geq1} \big(1-\prod_{i=1}^NQ_{1,i}^nQ_{2,i}^n\big)}
			\cdot \prod_{1\leq i<k\leq N} \tilde{f}_{k,i,-1}
			\cdot\prod_{i,k=1}^N
			\prod_{j=0}^{\infty} \tilde{f}_{i,k,j}
		\end{align}
		where $M(z;q)=\prod_{n=1}^\infty (1-zq^{-n})^n$, and
		\begin{align*}
			\tilde{f}_{i,k,j}=&\frac{M(Q_{1,i}Q_{1,k}\tilde{a}_{i,k,j};q)
				\cdot M(\tilde{a}_{i,k,j};q)}
			{M(Q_{1,k} \tilde{a}_{i,k,j};q)
				\cdot M(Q_{1,i} \tilde{a}_{i,k,j};q)},\\
			\tilde{a}_{i,k,j}=&
			Q_{2,k} \cdot \prod_{n=1}^{i-1}Q_{1,n}Q_{2,n}
			\cdot \prod_{n=k+1}^NQ_{1,n}Q_{2,n}
			\cdot \prod_{n=1}^N(Q_{1,n}Q_{2,n})^{j}.
		\end{align*}
	\end{prop}
	{\bf Proof:}
	By the rotation symmetry \eqref{eqn:rotation symm C} and the vacuum expectation value formula \eqref{eqn:clm0=<>} of topological vertex,
	we have
	\begin{align*}
		\tilde{Z}_N=\sum_{\mu^1,\nu^1,...,\mu^N,\nu^N}
		\prod_{i=1}^N(-Q_{1,i})^{|\mu^i|}(-Q_{2,i})^{|\nu^i|}
		\langle\mu^i|
		\Gamma_-(q^{\rho})\Gamma_+(q^{\rho})
		|\nu^{i}\rangle
		\cdot
		\langle\mu^{i+1,t}|
		\Gamma_-(q^{\rho})\Gamma_+(q^{\rho})
		|\nu^{i,t}\rangle.
	\end{align*}
	Similar to the proof of Proposition \ref{pro:Z_N as inf prod},
	we take the $w$-transform to the last factor in the above equation and then take a dual and shift the index $i\mapsto i-1$.
	\begin{align*}
		\tilde{Z}_N
		=\sum_{\mu^1,\nu^1,...,\mu^N,\nu^N \in\mathcal{P}}
		\prod_{i=1}^N&(-Q_{1,i})^{|\mu^i|}(-Q_{2,i})^{|\nu^i|}\\
		&\cdot
		\langle\mu^i|
		\Gamma_-(q^{\rho})\Gamma_+(q^{\rho})
		|\nu^{i}\rangle
		\cdot
		\langle\nu^{i-1}|
		\Gamma_-(-q^{\rho})^{-1}\Gamma_+(-q^{\rho})^{-1}
		|\mu^{i}\rangle.
	\end{align*}
	After summing over all $\mu^1,\mu^2,...,\mu^N$ and $\nu^{1},\nu^{2},...\nu^{N-1}$,
	we have
	\begin{align}\label{eqn:tZ_N last step}
		\begin{split}
			\tilde{Z}_N
			=&\sum_{\nu^1,...,\nu^N \in\mathcal{P}}
			\prod_{i=1}^N
			\langle\nu^{i-1}|
			\Gamma_-(-q^{\rho})^{-1}\Gamma_+(-q^{\rho})^{-1}
			(-Q_{1,i})^{L_0}
			\Gamma_-(q^{\rho})\Gamma_+(q^{\rho})
			(-Q_{2,i})^{L_0}|\nu^{i}\rangle\\
			=&\prod_{i=1}^N M(Q_{1,i};q)
			\cdot \prod_{1\leq i<k\leq N}
			\frac{M\big(Q_{1,k}\prod_{j=i}^{k-1}Q_{1,j}Q_{2,j};q\big)
				\cdot M\big(Q_{2,i}\prod_{j=i+1}^{k-1}Q_{1,j}Q_{2,j};q\big)}
			{M\big(\prod_{j=i}^{k-1}Q_{1,j}Q_{2,j};q\big)
				\cdot M\big(Q_{2,i}Q_{1,k}\prod_{j=i+1}^{k-1}Q_{1,j}Q_{2,j};q\big)}\\
			&\cdot\sum_{\nu^N\in\mathcal{P}}
			\langle \nu^N|
			\prod_{i=1}^N \Big(
			\Gamma_-(-\prod_{j=1}^{i-1}Q_{1,j}Q_{2,j}
			\cdot q^{\rho})^{-1}
			\Gamma_-(-Q_{1,i}\prod_{j=1}^{i-1}Q_{1,i}Q_{2,i}
			\cdot q^{\rho})\Big)\\
			&\cdot \bigg(\prod_{i=1}^N Q_{1,i}Q_{2,i}\bigg)^{{L_0}}
			\prod_{i=1}^N 
			\Big(\Gamma_+(-\prod_{j=i}^NQ_{1,j}Q_{2,j}
			\cdot q^{\rho})^{-1}
			\Gamma_+(-Q_{2,i}\prod_{j=i+1}^NQ_{1,j}Q_{2,j}
			\cdot q^{\rho})\Big)
			|\nu^N\rangle.
		\end{split}
	\end{align}
	
	Thus, we can apply the Lemma \ref{lem:mu,g--Qg++,mu} for $Q=\prod_{j=1}^N Q_{1,j}Q_{2,j}$ and
	\begin{align*}
		&x_{1,i}=-\prod_{j=1}^{i-1}Q_{1,j}Q_{2,j},
		\quad x_{2,i}=-Q_{1,i}\prod_{j=1}^{i-1}Q_{1,j}Q_{2,j}\\
		&y_{1,i}=-\prod_{j=i}^NQ_{1,j}Q_{2,j},
		\quad y_{2,i}=-Q_{2,i}\prod_{j=i+1}^NQ_{1,j}Q_{2,j},
		\quad 1\leq i \leq N
	\end{align*}
	to calculate the last term in equation \eqref{eqn:tZ_N last step}.
	This proposition then follows from the definitions of $\tilde{f}_{i,k,j}$ and $\tilde{a}_{i,k,j}$.
	$\Box$

	\begin{prop}
		For any positive integer $N$,
		we have
		\begin{align}
			\begin{split}
				\tilde{Z}_N
				=&\prod_{i=1}^N M(Q_{1,i};q)
				\cdot\sum_{\nu^1,...,\nu^N \in\mathcal{P}}
				\prod_{i=1}^N (-Q_{2,i})^{|\nu^i|}
				\cdot q^{-\|\nu^{i-1,t}\|^2/2-\|\nu^{i}\|^2/2}\\
				&\quad\quad\quad\cdot
				\prod_{(j,k)\in\nu^i}
				\frac{(1-Q_{1,i+1}q^{\nu^{i}_j+\nu^{i+1,t}_k-j-k+1})
					(1-Q_{1,i}q^{-\nu^{i-1,t}_k-\nu^i_j+j+k-1})}{(1-q^{-h(j,k)})^2}.
			\end{split}
		\end{align}
	\end{prop}
	{\bf Proof:}
	By the Schur expansion \eqref{eqn:def C} of the topological vertex,
	we have
	\begin{align*}
		\tilde{Z}_N
		=&\sum_{\mu^1,\nu^1,...,\mu^N,\nu^N \in\mathcal{P}}
		\prod_{i=1}^N(-Q_{1,i})^{|\mu^i|}(-Q_{2,i})^{|\nu^i|}
		s_{\nu^{i,t}}(q^\rho) s_{\mu^{i,t}}(q^{\nu^{i,t}+\rho})
		\cdot s_{\nu^{i}}(q^\rho) s_{\mu^{i+1}}(q^{\nu^{i}+\rho}).
	\end{align*}
	Thus,
	by taking an index shift $i\mapsto i-1$ to the last two terms in the above equation and taking Lemma \ref{lem:inf/inf=finite} to take the summation over $\mu^i$,
	we have
	\begin{align*}
		\tilde{Z}_N=&\prod_{i=1}^N M(Q_{1,i};q)
		\cdot\sum_{\nu^1,...,\nu^N \in\mathcal{P}}
		\prod_{i=1}^N (-Q_{2,i})^{|\nu^i|}
		s_{\nu^{i,t}}(q^\rho) s_{\nu^{i-1}}(q^\rho)\\
		&\cdot\prod_{(j,k)\in\nu^{i-1}}
		(1-Q_{1,i}q^{\nu^{i-1}_j+\nu^{i,t}_k-j-k+1})
		\cdot \prod_{(j,k)\in\nu^i}
		(1-Q_{1,i}q^{-\nu^{i-1,t}_k-\nu^i_j+j+k-1}).
	\end{align*}
	Thus,
	this proposition follows from the evaluation formula \eqref{eqn:schur q^rho} for the Schur functions and the following
	\begin{align*}
		\tilde{Z}_N=&\prod_{i=1}^N M(Q_{1,i};q)
		\cdot\sum_{\nu^1,...,\nu^N \in\mathcal{P}}
		\prod_{i=1}^N (-Q_{2,i})^{|\nu^i|}
		\cdot q^{-\|\nu^{i-1,t}\|^2/2-\|\nu^{i}\|^2/2}\\
		&\cdot\prod_{(j,k)\in\nu^{i-1}}
		\frac{(1-Q_{1,i}q^{\nu^{i-1}_j+\nu^{i,t}_k-j-k+1})}{(1-q^{-h(j,k)})}
		\cdot \prod_{(j,k)\in\nu^i}
		\frac{(1-Q_{1,i}q^{-\nu^{i-1,t}_k-\nu^i_j+j+k-1})}{(1-q^{-h(j,k)})}\\
		=&\prod_{i=1}^N M(Q_{1,i};q)
		\cdot\sum_{\nu^1,...,\nu^N \in\mathcal{P}}
		\prod_{i=1}^N (-Q_{2,i})^{|\nu^i|}
		\cdot q^{-\|\nu^{i-1,t}\|^2/2-\|\nu^{i}\|^2/2}\\
		&\cdot
		\prod_{(j,k)\in\nu^i}
		\frac{(1-Q_{1,i+1}q^{\nu^{i}_j+\nu^{i+1,t}_k-j-k+1})
			(1-Q_{1,i}q^{-\nu^{i-1,t}_k-\nu^i_j+j+k-1})}{(1-q^{-h(j,k)})^2}.
	\end{align*}
	$\Box$

	\begin{prop}[=Proposition \ref{cor:main2}]
		We have
		\begin{small}
			\begin{align}\label{eqn:main2 app}
				&\sum_{\nu^1,...,\nu^N \in\mathcal{P}}
				\prod_{i=1}^N (-s_i)^{|\nu^i|}
				\prod_{(j,k)\in\nu^{i}}
				\frac{\big(t_{i+1}+(\nu^{i}_j+\nu^{i+1,t}_k-j-k+1)\big)
					\big(t_i-(\nu^{i-1,t}_k+\nu^i_j-j-k+1)\big)}
				{h(j,k)^2} \nonumber\\
				&\quad\quad\quad\quad\quad\quad\quad\quad=
				\frac{1}
				{\prod\limits_{n\geq1}
					(1-s^n)}
				\cdot \prod\limits_{1\leq i<k\leq N} (1-\tilde{b}_{k,i,-1})^{t_it_k}
				\cdot\prod_{i,k=1}^N
				\prod_{j=0}^{\infty} (1-\tilde{b}_{i,k,j})^{t_it_k},
			\end{align}
		\end{small}
		where $t_{N+1}=t_1$, $s=\prod_{i=1}^Ns_i$ and
		\begin{align*}
			\tilde{b}_{i,k,j}=
			\prod_{n=1}^{i-1}s_n
			\cdot\prod_{n=k}^N s_n
			\cdot\prod_{n=1}^N (s_n)^j.
		\end{align*}
	\end{prop}
	{\bf Proof:}
	After the change of variables \eqref{eqn:change beta} and taking $\beta\rightarrow0$,
	the left-hand side of equation \eqref{eqn:main2} gives
	\begin{align*}
		\sum_{\nu^1,...,\nu^N}
		\prod_{i=1}^N (-s_i)^{|\nu^i|}
		\prod_{(j,k)\in\nu^i}
		\frac{\big(t_{i+1}+(\nu^{i}_j+\nu^{i+1,t}_k-j-k+1)\big)
			\big(t_i-(\nu^{i-1,t}_k+\nu^i_j-j-k+1)\big)}
		{h(j,k)^2}.
	\end{align*}
	
	For $\tilde{f}_{i,k,j}$,
	similar to the method in proving Corollary \ref{cor:main},
	we have
	\begin{align*}
		\tilde{f}_{i,k,j}\rightarrow&
		\frac{\prod_{n=1-t_i-t_k}^0 (1-\tilde{b}_{i,k,j})^{n+t_i+t_k}}
		{\prod_{n=1-t_k}^0 (1-\tilde{b}_{i,k,j})^{n+t_k}
			\cdot\prod_{n=1-t_i}^0 (1-\tilde{b}_{i,k,j})^{n+t_i}}\\
		=&(1-\tilde{b}_{i,k,j})^{t_it_k}.
	\end{align*}
	This proposition is thus proved by directly applying the limit $\beta\mapsto0$ to equation \eqref{eqn:main2}.
	$\Box$

	When $N=1$,
	the equation \eqref{eqn:main2 app} reduces to the original Nekrasov--Okounkov formula in \cite{NO06} (see also \cite{CO12,H08,W06} and the derivation in Section 2 of \cite{INRS}):
	\begin{align*}
		&\sum_{\nu\in\mathcal{P}}
		(-s)^{|\nu|}
		\cdot
		\prod_{(j,k)\in\nu}
		\frac{\big(t+h(j,k)\big)\big(t-h(j,k)\big)}
		{h(j,k)^2}
		=\prod\limits_{n\geq1} (1-s^n)^{t^2-1}.
	\end{align*}

	\section{Conflict of interest and data availability statement}
	The author states that there is no conflict of interest, and
	no datasets were generated or analysed during the current study.

	\vspace{.2in}
	{\em Acknowledgements}.
	The author would like to thank Xiaobo Liu, Xinxing Tang, Zhiyuan Wang, Chunlei Liu and Xuhui Zhang for helpful discussions.
	The author is supported by the NSFC grants (No. 12288201, 12401079),
	the China Postdoctoral Science Foundation (No. 2023M743717)
	and China National Postdoctoral Program for Innovative Talents (No. BX20240407).
	\vspace{.2in}

	\renewcommand{\refname}{Reference}
	\bibliographystyle{plain}
	\bibliography{reference}
	\vspace{30pt} \noindent
\end{document}